\documentclass[lettersize,journal]{IEEEtran}
\usepackage{amssymb}
\usepackage{algorithm}
\usepackage{algorithmic}
\usepackage{amsmath,amsfonts}
\usepackage{amsthm}
\usepackage{array}
\usepackage{arydshln}
\usepackage{bbding}
\usepackage{bigints}
\usepackage{booktabs}
\usepackage[caption=false,font=normalsize,labelfont=sf,textfont=sf]{subfig}
\usepackage{cite}
\usepackage{color}
\usepackage{diagbox}
\usepackage{epsfig,latexsym}
\usepackage{epstopdf}
\usepackage{graphicx}
\usepackage{fancyhdr}
\usepackage{float}
\usepackage{flushend}
\usepackage{indentfirst}
\usepackage{lastpage}
\usepackage{makecell}
\usepackage{mathtools}
\usepackage{multirow}
\usepackage{pifont}
\usepackage{psfrag}
\usepackage{setspace}
\usepackage{stfloats}
\usepackage{subfig}
\usepackage{subfloat}
\usepackage{textcomp}
\usepackage{times}
\usepackage{verbatim}
\usepackage{url}
\usepackage{hyperref}
\usepackage{balance}
\usepackage{cleveref}
\newcommand{\cmark}{\ding{52}}

\theoremstyle{remark}
\newtheorem{theorem}{\quad \textbf{Theorem}}

\newtheorem{corollary}{\quad \textbf{Corollary}}

\allowdisplaybreaks[4]

\begin{document}
\title{Holographic Metasurface-Based Beamforming for Multi-Altitude LEO Satellite Networks}
\author{Qingchao Li, \textit{Member, IEEE},
Mohammed El-Hajjar, \textit{Senior Member, IEEE},
Kaijun Cao,\\
Chao Xu, \textit{Senior Member, IEEE},
Harald Haas, \textit{Fellow, IEEE},
and Lajos Hanzo, \textit{Life Fellow, IEEE}

\thanks{This work was supported by the Future Telecoms Research Hub, Platform for Driving Ultimate Connectivity (TITAN),
sponsored by the Department of Science Innovation and Technology (DSIT) and the Engineering and Physical Sciences Research Council (EPSRC) under Grant EP/X04047X/1 and Grant EP/Y037243/1. M. El-Hajjar would like to acknowledge the financial support of the Engineering and Physical Sciences Research Council (EPSRC) projects under grant EP/X04047X/2, EP/X04047X/1 and EP/Y037243/1. L. Hanzo would also like to acknowledge the financial support of the Engineering and Physical Sciences Research Council (EPSRC) projects under grant EP/W016605/1, EP/X01228X/1, EP/Y026721/1, EP/W032635/1, as well as of the European Research Council's Advanced Fellow Grant QuantCom (Grant No. 789028). \textit{(Corresponding author: Lajos Hanzo.)}

Qingchao Li, Mohammed El-Hajjar, Chao Xu and Lajos Hanzo are with the School of Electronics and Computer Science, University of Southampton, Southampton SO17 1BJ, U.K. (e-mail: qingchao.li@soton.ac.uk; meh@ecs.soton.ac.uk; cx1g08@ecs.soton.ac.uk; lh@ecs.soton.ac.uk).

Kaijun Cao is with the School of Mathematical Sciences, University of Southampton, Southampton SO17 1BJ, U.K. (e-mail: kc8g22@soton.ac.uk).

Harald Haas is with the Department of Engineering, Electrical Engineering Division, Cambridge University, CB3 0FA Cambridge, U.K. (e-mail: huh21@cam.ac.uk).}
}

\maketitle

\begin{abstract}
Low Earth Orbit (LEO) satellite networks are capable of improving the global Internet service coverage. In this context, we propose a hybrid beamforming design for holographic metasurface based terrestrial users in multi-altitude LEO satellite networks. Firstly, the holographic beamformer is optimized by maximizing the downlink channel gain from the serving satellite to the terrestrial user. Then, the digital beamformer is designed by conceiving a minimum mean square error (MMSE) based detection algorithm for mitigating the interference arriving from other satellites. To dispense with excessive overhead of full channel state information (CSI) acquisition of all satellites, we propose a low-complexity MMSE beamforming algorithm that only relies on the distribution of the LEO satellite constellation harnessing stochastic geometry, which can achieve comparable throughput to that of the algorithm based on the full CSI in the case of a dense LEO satellite deployment. Furthermore, it outperforms the maximum ratio combining (MRC) algorithm, thanks to its inter-satellite interference mitigation capacity. The simulation results show that our proposed holographic metasurface based hybrid beamforming architecture is capable of outperforming the state-of-the-art antenna array architecture in terms of its throughput, given the same physical size of the transceivers. Moreover, we demonstrate that the beamforming performance attained can be substantially improved by taking into account the mutual coupling effect, imposed by the dense placement of the holographic metasurface elements.
\end{abstract}

\section{Introduction}
\IEEEPARstart{A}{lthough} traditional terrestrial communication networks have been widely rolled out across the globe for providing significantly increased throughput by leveraging a whole suite of sophisticated techniques, it is still challenging to support global Internet connectivity~\cite{tsai2024distributionally}, \cite{xiang2024massive}. The growing interest in satellite constellations harnessed for ubiquitous communications and global Internet coverage highlights the importance of low Earth Orbit (LEO) satellites in bridging the digital divide and enhancing connectivity in remote regions~\cite{cao2022edge}, \cite{wang2024ultra}. Compared to its Geostationary Earth Orbit (GEO) and Medium Earth Orbit (MEO) counterparts, LEO satellites offer notable advantages such as reduced latency and higher data rates due to their proximity to the Earth~\cite{zeng2024multi}, \cite{jiang2020reinforcement}.

\subsection{Related Work}
Substantial research efforts have also been dedicated to the exploration of various aspects in LEO satellite communications, including multiple-input and multiple-output (MIMO) technologies, hybrid beamforming methods, and robust secure transmission strategies. Specifically, Li \textit{et al.}~\cite{li2021downlink} focused their attention on the design of downlink transmission strategies for massive MIMO LEO satellite communications. It was demonstrated that using massive MIMO is beneficial for enhancing the spectral efficiency and mitigating the limitations imposed by the dynamic nature of LEO satellites. In this context, the advanced beamforming designs have been conceived for robust communications between satellites and the base stations (BSs) on the ground. Simulation results indicated significant throughput improvements, demonstrating the benefits of harnessing massive MIMO in LEO satellite systems. To avoid the excessive complexity and energy consumption of the fully digital beamforming architecture, You \textit{et al.}~\cite{you2022hybrid} conceived hybrid analog-digital precoding techniques for LEO satellite systems. Specifically, a novel hybrid precoding architecture was proposed relying on statistical channel state information (CSI) for maximizing the energy efficiency and/or reducing the complexity. Their simulation results demonstrated that the proposed hybrid precoding schemes achieve significant energy efficiency gains over existing baselines, especially when discrete phase shift networks are employed for analog precoding. Considering the imperfect hardware factors, including low resolution phase shifters and nonlinear power amplifiers (NPAs), an efficient algorithmic approach was formulated in~\cite{you2022massive} for LEO satellite networks in the context of a twin-resolution phase shifting (TRPS) based hybrid precoding problem. This design struck an attractive energy efficiency versus computational complexity trade-off. In~\cite{huang2023qos}, Huang \textit{et al.} presented a quality of service (QoS)-aware precoding design, which aimed for optimizing both the energy efficiency and user satisfaction in the downlink of massive MIMO aided LEO satellite communications. A multi-objective optimization problem was formulated by striking a trade-off between the energy efficiency and the proportion of users meeting their QoS requirements. Furthermore, an efficient algorithm was developed for solving a multi-objective problem, resulting in improved QoS and energy performance. In~\cite{liu2022robust}, Liu \textit{et al.} focused on robust downlink precoding strategies for LEO satellite systems operating under per-antenna power constraints. Moreover, a robust precoding scheme was proposed for optimizing the performance under individual power limits for each antenna. The proposed technique enhanced the communications integrity, particularly in environments associated with fluctuating channel conditions and stringent power constraints.

Albeit the massive MIMO technique is of substantial benefit for high-speed data services and global connectivity in LEO satellite networks, traditional antenna arrays face limitations in terms of scalability, cost, and power consumption. It is infeasible to realize extremely large-scale MIMO (XL-MIMO) schemes relying on a large number of conventional radio frequency (RF) chains and active antennas due to the excessive power consumption requirements~\cite{huang2020holographic}, \cite{gong2023holographic}. As a remedy, holographic MIMO (HMIMO) technology has emerged as a promising design alternative, exhibiting improved hardware efficiency and energy efficiency. This ambitious objective is achieved by utilizing a spatially near-continuous aperture and holographic radios having reduced power consumption and fabrication cost~\cite{li2022reconfigurable_iot}, \cite{li2023reconfigurable}, \cite{yoo2023sub}, \cite{deng2023reconfigurable}. Recent advances in non-terrestrial networks highlighted the effectiveness of metasurface-based receiver architectures, which are capable of enhancing the communication quality and mitigating interference through configurable multi-layer or multi-functional metasurface designs~\cite{sun2024multi}, \cite{an2024exploiting}, \cite{sun2024active}. In~\cite{deng2022holographic}, Deng \textit{et al.} introduces a reconfigurable holographic surface (RHS) aided uplink communication system, where a user terminal equipped with RHS transmits data to multiple LEO satellites. A novel holographic beamforming algorithm was proposed for maximizing the sum rate, which is proved to be robust against tracking errors in satellite positions. Their simulation results indicated that the RHS outperforms traditional phased arrays in terms of its sum-rate and cost-efficiency, owing to its compact element spacing and low hardware costs. Furthermore, in~\cite{hu2023holographic}, a closed-form expression was derived for maximizing the sum rate of LEO satellite communications relying on a RHS. The authors theoretically analyzed the minimum number of RHS elements required for the sum-rate of the RHS-aided system to exceed that of the phased array system. The simulation results showed that the RHS-based LEO satellite communications is capable of outperforming traditional phased array based systems in terms of both the sum rate and hardware efficiency. Stacked intelligent metasurfaces (SIM) are also capable of improving the performance of LEO satellite communications. Lin \textit{et al.}~\cite{lin2024stacked} proposed a SIM-based multi-beam LEO system, which performs downlink precoding in the wave domain for reducing both the processing latency and computational burden. Based on the statistical CSI, an optimization problem was formulated for maximizing the ergodic sum rate, which was solved by a customized alternating optimization algorithm. The results demonstrated significant improvements in sum rate and computational efficiency compared to traditional digital precoding methods.

The above contributions were focused on single LEO satellite architectures. Considering the ultra-dense deployment of LEO satellites to achieve global connectivity, several research efforts were dedicated to LEO constellations, resulting in inter-satellite interference. In~\cite{okati2020downlink}, Okati \textit{et al.} derived the analytically tractable expression of the downlink coverage probability and the data rate of LEO satellite constellations relying on stochastic geometry. Jung \textit{et al.}~\cite{jung2022performance} focused on the performance analysis of LEO satellite communication systems under the shadowed-Rician fading model. The binomial point process (BPP) was employed to model the distribution of LEO satellites. Based on this, both the outage probability and the system throughput were evaluated. In~\cite{park2022tractable}, Park \textit{et al.} presented a tractable method for the downlink coverage analysis, highlighting the pivotal role of satellite density and altitude in optimizing the network performance attained. This approach facilitates the efficient characterization of diverse deployment scenarios. Moreover, Sun \textit{et al.} in~\cite{sun2023fine} utilized a homogeneous Poisson Point Process (PPP) to analyze LEO networks, focusing on user fairness and transmission reliability. It showed that deploying satellites at lower altitudes benefits dense networks by enhancing both the coverage probability and user fairness, while higher altitudes are preferable for sparser networks. In~\cite{choi2024modeling}, Choi \textit{et al.} provided a comprehensive analysis of downlink communications in heterogeneous LEO satellite networks using Cox point processes to model the distribution of satellites. They characterized both closed and open access scenarios, demonstrating that open access significantly improves the coverage probability. Complementing the homogeneous models, Okati \textit{et al.} in~\cite{okati2022nonhomogeneous} applied non-homogeneous stochastic geometry for analyzing massive LEO satellite constellations. Their approach accounts for the variations in satellite density and spatial distribution across different regions. Then various new performance metrics, such as the conditional coverage probability and user throughput, were derived under non-homogeneous conditions. The results demonstrated that considering non-homogeneous distributions provides a more accurate representation of real-world satellite networks. Furthermore, Hu \textit{et al.}~\cite{hu2024performance} investigated the end-to-end performance of LEO satellite-aided shore-to-ship communications using stochastic geometry in maritime communication contexts. They evaluated the impact of influential factors such as satellite altitude, transmission power and environmental conditions on the communication links between ports and ships.

To meet the growing demand of beamless global connectivity and efficient communication infrastructure, multi-altitude LEO satellite networks have emerged as a pivotal solution. They can offer numerous advantages in terms of coverage, latency, and bandwidth efficiency over the single-altitude LEO satellite architecture in~\cite{okati2020downlink,jung2022performance,
park2022tractable,sun2023fine,choi2024modeling,okati2022nonhomogeneous,hu2024performance}. In particular, Okati \textit{et al.}~\cite{okati2023stochastic} provided a comprehensive analysis of the coverage probabilities in multi-altitude LEO satellite networks, where the satellites are modelled as a BPP assuming their altitude is an arbitrarily distributed random variable. Their simulation results showed that the coverage performance becomes saturated when the constellation size reaches a certain threshold. In~\cite{choi2024cox}, Choi \textit{et al.} introduced an innovative technique of modeling satellite networks using Cox point processes. Specifically, the orbits vary in altitude and the distribution of satellites on each orbit are modelled as a linear PPP. Some useful statistics, including the distribution of the distance from the typical terrestrial user to its nearest visible satellite and the outage probability, were theoretically derived.

\begin{table*}
\footnotesize
\begin{center}
\caption{Contrasting the novelty of our paper to the existing LEO satellite communications literature~\cite{li2021downlink}, \cite{you2022hybrid}, \cite{you2022massive}, \cite{huang2023qos}, \cite{liu2022robust}, \cite{deng2022holographic}, \cite{hu2023holographic}, \cite{lin2024stacked}, \cite{okati2020downlink}, \cite{jung2022performance}, \cite{park2022tractable}, \cite{sun2023fine}, \cite{choi2024modeling}, \cite{okati2022nonhomogeneous}, \cite{hu2024performance}, \cite{okati2023stochastic}, \cite{choi2024cox}.}
\label{Table_literature}
\begin{tabular}{|c|c|c|c|c|c|}
\hline
     & \makecell[c]{Our paper} & \cite{li2021downlink}, \cite{you2022hybrid}, \cite{you2022massive}, \cite{huang2023qos} & \cite{deng2022holographic}, \cite{hu2023holographic}, \cite{lin2024stacked} & \cite{okati2020downlink}, \cite{jung2022performance}, \cite{park2022tractable}, \cite{sun2023fine}, \cite{choi2024modeling}, \cite{okati2022nonhomogeneous}, \cite{hu2024performance} & \cite{okati2023stochastic}, \cite{choi2024cox} \\
\hline
    Beamforming design & \makecell[c]{\cmark} & \makecell[c]{\cmark} & \makecell[c]{\cmark} &  & \\
\hdashline
    Holographic metasurface & \makecell[c]{\cmark} &  & \makecell[c]{\cmark} &  & \\
\hdashline
    LEO satellite constellation & \makecell[c]{\cmark} &  &  & \makecell[c]{\cmark} & \makecell[c]{\cmark} \\
\hdashline
    Multi-altitude satellites & \makecell[c]{\cmark} &  &  &  & \makecell[c]{\cmark} \\
\hdashline
    Inter-satellite interference mitigation & \makecell[c]{\cmark} &  &  &  & \\
\hline
\end{tabular}
\end{center}
\end{table*}

\subsection{Motivation}
The above LEO satellite communication architectures have the following limitations. Firstly, the existing beamforming designs conceived for multi-satellite networks ignore the inter-satellite interference, which significantly limits the data rate in dense satellite constellations. Although inter-satellite interference can be mitigated through cooperative satellite networks, as suggested in~\cite{zhang2024multi}, which brings in increased cost of establishing and maintaining inter-satellite links. Secondly, the above beamforming designs rely on the acquisition of full CSI, which significantly increases the communication overhead. The process of obtaining full CSI across all satellite links not only requires substantial bandwidth but also leads to high computational complexity, especially in dense satellite constellations. To deal with these challenges, we propose a holographic metasurface-based beamforming architecture for multi-altitude LEO satellite networks, purely relying on the statistical distribution of the LEO satellite constellation. Against this background, Table~\ref{Table_literature} explicitly contrasts our contributions to the literature at a glance, which are further detailed as follows.

\begin{itemize}
  \item The holographic metasurfaces achieve high directional gain despite using a compact antenna for mitigating the severe path loss of satellite communications. In this paper, we conceive a metasurface based hybrid holographic and digital beamforming architecture for the multi-altitude LEO satellite downlink, where a holographic metasurface is employed by the terrestrial user for spectral-efficient information transfer.
  \item Since designing the metasurface coefficients of the holographic beamformer and of the digital receiver combining (RC) vector is a non-convex problem, we decompose it into two sub-problems. Specifically, the RF holographic beamformer is optimized for maximizing the channel gain from the serving satellite to the terrestrial user. Once the holographic beamformer weights are given, the baseband equivalent channel from satellites to the RF chains can be obtained. Afterwards, the digital RC vector is optimized based on the minimum mean square error (MMSE) criterion for mitigating the interference imposed by the interfering satellites on the terrestrial user.
  \item To avoid the high overhead of acquiring full CSI, we propose a low-complexity MMSE detection algorithm. In this approach, the digital beamformer is designed based on the statistical characteristics of the LEO satellite constellation, specifically leveraging the average number and spatial distribution of interfering satellites within the visible region through stochastic geometry. By focusing on the distributional properties of visible interfering satellites, this method significantly reduces channel estimation complexity.
  \item Our numerical results show that the proposed holographic metasurface based hybrid beamforming architecture is capable of achieving higher throughput than the state-of-the-art (SoA) antenna array architecture. More explicitly, they show that the MMSE RC algorithm outperforms the maximum ratio combining (MRC) algorithm, thanks to the inter-satellite interference mitigation. The throughput can be improved by explicitly considering the mutual coupling effect in the beamforming design, which arises due to the dense placement of the holographic metasurface elements. Furthermore, the proposed low-complexity MMSE algorithm based on the distribution of the satellites can achieve similar throughput to that of the idealized algorithm based on the perfect knowledge of the full CSI associated with all satellites in the case of a dense deployment of LEO satellites.
\end{itemize}

\subsection{Organization}
The rest of this paper is organized as follows. In Section~\ref{System_Model}, we present the system model, while both the hybrid holographic and the digital beamforming design is described in Section~\ref{Beamforming_Design}. Our simulation results are presented in Section~\ref{Simulation_Results}, while we conclude in Section~\ref{Conclusion}.

\subsection{Notations}
Vectors and matrices are denoted by boldface lower and upper case variables, respectively; sets are denoted by calligraphic letters; $(\cdot)^{\text{T}}$ and $(\cdot)^{\text{H}}$ represent the operation of transpose and Hermitian transpose, respectively; $|a|$ and $\angle a$ denote the amplitude and angle of the complex scalar $a$, respectively; $|\mathcal{A}|$ represents the cardinality of the set $\mathcal{A}$; $\|\mathbf{a}\|$ denotes the norm of the vector $\mathbf{a}$; $\mathbb{C}^{m\times n}$ is the space of $m\times n$ complex-valued matrices; $\mathbf{0}_{N}$ is the $N\times1$ zero vector; $\mathbf{I}_{N}$ represents the $N\times N$ identity matrix; $\mathbf{Diag}\{a_1,a_2,\cdots,a_N\}$ denotes a diagonal matrix having elements of $a_1,a_2,\cdots,a_N$ in order; $a_n$ is the $n$th element in the vector $\mathbf{a}$; $\mathcal{CN}(\boldsymbol{\mu},\mathbf{\Sigma})$ is a circularly symmetric complex Gaussian random vector with the mean $\boldsymbol{\mu}$ and covariance matrix $\mathbf{\Sigma}$; $f_X(x)$ and $F_X(x)$ represent the probability density function (PDF) and the cumulative distribution function (CDF) of the random variable $X$, respectively.

\section{System Model}\label{System_Model}
\begin{figure}[!t]
    \centering
    \includegraphics[width=2.7in]{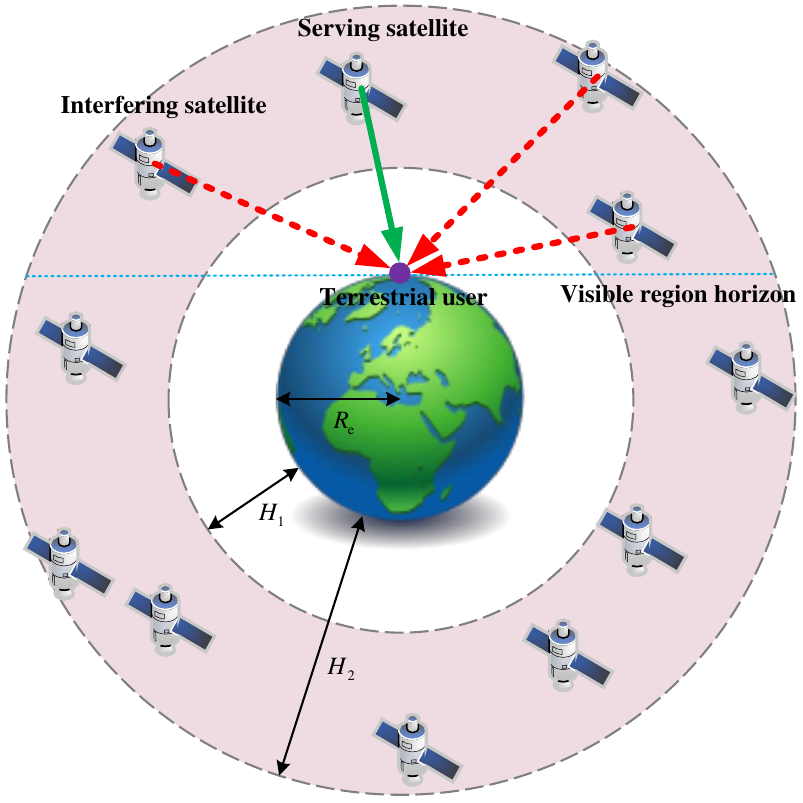}
    \caption{System model of holographic metasurface-based multi-altitude LEO satellite networks.}\label{Fig_system_model}
\end{figure}

In this section, we describe our proposed holographic metasurface-based multi-altitude LEO satellite network. The system model of the downlink multi-altitude LEO satellite network is shown in Fig.~\ref{Fig_system_model}\footnote{In this paper, we investigate narrowband satellite communication networks. In wideband networks, both spatial-wideband effects and frequency-selective effects should be considered~\cite{xu2024near}. The satellite communications of wideband networks considering the spatial-wideband effect and the frequency-selective effect is part of our future work.}. In contrast to the conventional single-altitude LEO satellite networks, which consider a constellation consisting of LEO satellites at the same altitude, the multi-altitude LEO satellite network considers a constellation consisting of LEO satellites at different altitudes, positioned between $H_1$ and $H_2$. We assume that the Earth is a perfect sphere with a radius of $R_\mathrm{e}$ centred at the origin of $(0,0,0)\in\mathbb{R}^3$ in the three-dimensional (3D) Cartesian coordinate system. We also assume that the single-antenna LEO satellites are uniformly distributed within the shell determined by $\Omega=\{R_\mathrm{e}+H_1\leq\sqrt{x^2+y^2+z^2}\leq R_\mathrm{e}+H_2|(x,y,z)\in\mathbb{R}^3\}$ forming a 3D BPP~\cite{jung2022performance}, denoted as $\mathcal{A}$, ensuring an even spatial distribution across the different altitudes in the specified range. Furthermore, we denote the number of satellites within the region of $\Omega$ as $|\mathcal{A}|$.

Each terrestrial user is served by the nearest satellite, which is referred to as the serving satellite, resulting in a spherical Voronoi tessellation for the satellites' coverage areas. We assume that all satellites in the region of $\Omega$ share the same time/frequency resource. Hence, a typical terrestrial terminal served by the nearest satellite is interfered by all other satellites. We consider a typical terrestrial user located at the Cartesian coordinate of $(0,0,R_\mathrm{e})$. We denote the serving satellite as the 0th satellite, while the interfering satellites are indexed as the 1st, 2nd, $\cdots$, $(|\mathcal{A}|-1)$st satellites, based on their distances from the terrestrial user considered. Furthermore, we denote the coordinates of these interfering satellites as $\mathbf{p}_1,\mathbf{p}_2,\cdots,\mathbf{p}_{|\mathcal{A}|-1}$, respectively. As illustrated in Fig.~\ref{Fig_system_model}, all the satellites that are visible above the horizon can communicate with the terrestrial user. The visible region can be represented as $\Omega'=\{z\geq R_\mathrm{e}|(x,y,z)\in\Omega\}$.

\subsection{Holographic Metasurface-Based Beamforming}
Due to the large distance from the satellites to the terrestrial users, a holographic metasurface is employed by each terrestrial user to compensate the signal attenuation due to the path loss. The complete architecture includes a holographic beamformer and a digital beamforming.

\begin{figure}[!t]
    \centering
    \includegraphics[width=2.7in]{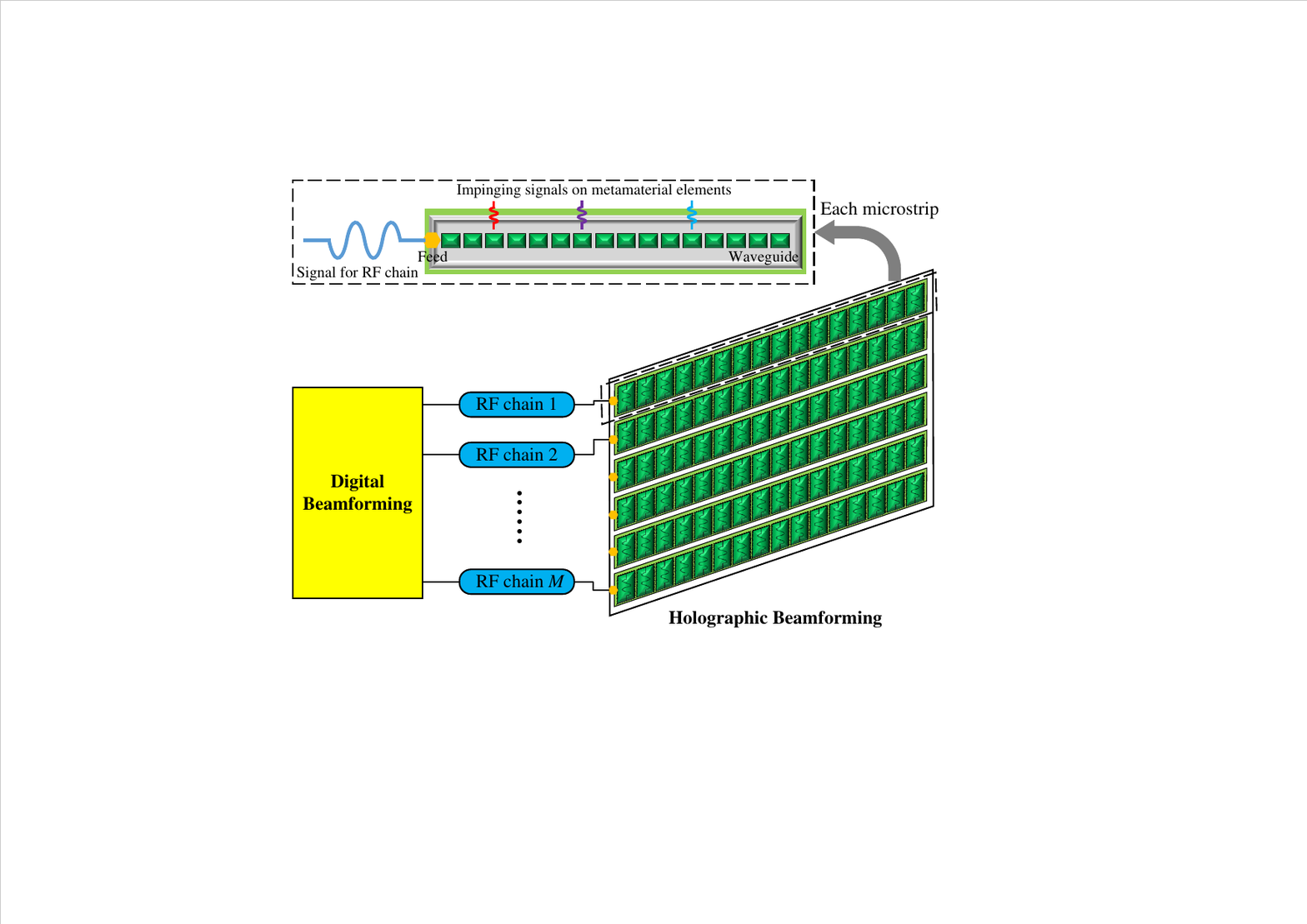}
    \caption{Holographic metasurface-based hybrid beamforming architecture.}\label{Fig_HMA}
\end{figure}

As shown in Fig.~\ref{Fig_HMA}, at the holographic metasurface of the beamformer is composed of $M$ microstrips, each of which is connected to an RF-chain. Each microstrip consists of three components, including a feed, a waveguide and $N$ sub-wavelength metamaterial elements. Specifically, each element in the microstrip is made of artificial composite material, which is capable of adjusting the coefficients of the electromagnetic (EM) waves with the aid of a software controller, such as a field programmable gate array (FPGA)~\cite{an2023stacked}, \cite{an2023stacked_arxiv}, \cite{li2024stacked}, \cite{li2024energy}, \cite{li2024performance}. The waveguide acts as the propagation medium of the EM wave spanning from the reconfigurable metasurface elements to the feed. The feed then transforms the EM wave into high frequency current for the RF-chain. Afterwards, the RF-chain converts the RF signals to the baseband signals for the digital beamformer.

We denote the weighting coefficient of the $n$th reconfigurable metasurface element at the $m$th microstrip as $\beta_n^{(m)}$. Under the Lorentzian-constrained phase model of~\cite{xu2024near}, \cite{zhang2022beam}, \cite{li2023near}, we can get
\begin{align}\label{System_Model_1}
    \beta_n^{(m)}=\frac{\jmath+\mathrm{e}^{\jmath\phi_n^{(m)}}}{2},
    \ m=1,2,\cdots,M,\ n=1,2,\cdots,N,
\end{align}
where $\phi_n^{(m)}\in[0,2\pi)$. Furthermore, the response from the $n$th reconfigurable metasurface element at the $m$th microstrip to its connected feed is given by
\begin{align}\label{System_Model_2}
    q_n^{(m)}=\mathrm{e}^{-\jmath\frac{2\pi}{\lambda}\xi_n^{(m)}},
\end{align}
where $\xi_n^{(m)}$ is the distance between the corresponding reconfigurable metasurface element and its connected feed.

\subsection{Channel Model}
In this section, we describe the channel model between the satellites and the terrestrial user. We denote the link spanning from the $l$th satellite to the $m$th microstrip of the terrestrial user as $\mathbf{f}^{(l,m)}\in\mathbb{C}^{N\times1}$, given by
\begin{align}\label{System_Model_3}
    \mathbf{f}^{(l,m)}=\sqrt{\varrho_l}\mathbf{g}^{(l,m)}.
\end{align}
In (\ref{System_Model_3}), $\varrho_l$ represents the link attenuation between the $l$th satellite and the terrestrial user, given by
\begin{align}\label{System_Model_4}
    \varrho_l=\varsigma\zeta\left(\frac{\lambda}{4\pi}\right)^2D_l^{-\alpha},
\end{align}
where $\lambda$ denotes the carrier wavelength, $\varsigma$ is the rain attenuation coefficient, $\zeta$ represents the antenna gain, $\alpha$ denotes the path loss exponent and $D_l$ is the distance between the $l$th satellite and the terrestrial user. Furthermore, $\mathbf{g}^{(l,m)}$ represents the small-scale fading. Referring to~\cite{huang2020energy}, the shadowed-Rician fading model is widely employed for describing the distribution of the channel links in satellite communications. Specifically, the $n$th entry in $\mathbf{g}^{(l,m)}$, denoted as $g^{(l,m)}_n$, represents the small scale fading between the $l$th satellite and the $n$th reconfigurable metasurface element of the $m$th microstrip at the terrestrial user. We denote the shadowed-Rician fading as $\mathcal{SR}(\omega,b_0,\upsilon)$, where $\omega$ is the average power of the line-of-sight (LoS) component, $2b_0$ is the average power of the scattered component and $\upsilon$ is the Nakagami parameter. The CDF of the channel's power gain $|g^{(l,m)}_n|^2$ is given by
\begin{align}\label{System_Model_5}
    \notag &F_{|g^{(l,m)}_n|^2}(x)=\left(\frac{2b_0}{2b_0\upsilon+\omega}\right)^\upsilon\cdot\\
    &\sum_{i=0}^\infty\frac{(\upsilon)_i}{i!\Gamma(i+1)}
    \left(\frac{\omega}{2b_0\upsilon+\omega}\right)^i
    \Gamma_\mathrm{i}\left(i+1,\frac{1}{2b_0}x\right),
\end{align}
where $(\upsilon)_i$ is the Pochhammer symbol, $\Gamma(\cdot)$ is the Gamma function, and $\Gamma_\mathrm{i}(\cdot,\cdot)$ is the lower incomplete Gamma function.

For the mutual coupling, we adopt the Z-parameter based representation to model this effect~\cite{balanis2016antenna}\footnote{The mutual coupling model used here is based on the analysis in~\cite{balanis2016antenna} and has been widely applied in holographic metasurfaces, demonstrating its general applicability in dense antenna arrays~\cite{xu2024near}.} Specifically, the mutual coupling matrix, denoted as $\mathbf{C}\in\mathbb{C}^{MN\times MN}$, is
\begin{align}\label{System_Model_6}
    \mathbf{C}=\left(Z_A+Z_L\right)\left(\mathbf{Z}+Z_L\mathbf{I}\right)^{-1},
\end{align}
where $Z_A$ is the antenna impedance and $Z_L$ is the load impedance, both of which are fixed as 50 Ohms. Furthermore, $\mathbf{Z}\in\mathbb{C}^{MN\times MN}$ is the mutual impedance matrix, with the $(i_1,i_2)$th entry represented as
\begin{align}\label{System_Model_7}
    \mathbf{Z}_{i_1,i_2}=Z_A,
\end{align}
for $i_1=i_2$, and
\begin{align}\label{System_Model_8}
    \notag&\mathbf{Z}_{i_1,i_2}
    =60\mathcal{C}_\mathrm{I}\left(\frac{2\pi d_{i_1,i_2}}{\lambda}\right)
      -60\mathcal{S}_\mathrm{I}\left(\frac{2\pi d_{i_1,i_2}}{\lambda}\right)\\
    \notag&-30\mathcal{C}_\mathrm{I}
    \left(\frac{2\pi\left(\tilde{d}_{i_1,i_2}+\delta_0\right)}{\lambda}\right)
    +30\mathcal{S}_\mathrm{I}
    \left(\frac{2\pi\left(\tilde{d}_{i_1,i_2}+\delta_0\right)}{\lambda}\right)\\
    &-30\mathcal{C}_\mathrm{I}
    \left(\frac{2\pi\left(\tilde{d}_{i_1,i_2}-\delta_0\right)}{\lambda}\right)
    +30\mathcal{S}_\mathrm{I}
    \left(\frac{2\pi\left(\tilde{d}_{i_1,i_2}-\delta_0\right)}{\lambda}\right),
\end{align}
for $i_1\neq i_2$, where $\tilde{d}_{i_1,i_2}=\sqrt{d_{i_1,i_2}^2+\delta_0^2}$. In (\ref{System_Model_8}), $d_{i_1,i_2}$ represents the distance between the $i_1$th and the $i_2$th metasurface element, $\delta_0$ is the dipole length, $\mathcal{C}_\mathrm{I}$ denotes the cosine integral and $\mathcal{S}_\mathrm{I}$ represents the sine integral.

\subsection{Distribution of Satellite Constellation}
In this section, we theoretically derive the distribution of the distance between the satellites and the terrestrial user.

Firstly, we derive the CDF of any specific satellite being in the visible region $\Omega'$ of the terrestrial user. For ease of exposition, we define a spherical cap $\Omega_1'=\{x^2+y^2+z^2\leq (R_\mathrm{e}+H_1)^2,z\geq R_\mathrm{e}|(x,y,z)\in\mathbb{R}^3\}$. Then its volume $V_1'$ can be calculated as
\begin{align}\label{System_Model_8_1}
    V_1'=\frac{\pi}{3}H_1\left(2R_\mathrm{e}+H_1\right)\left(3R_\mathrm{e}+2H_1\right).
\end{align}
Similarly, we define a second spherical cap $\Omega_2'=\{x^2+y^2+z^2\leq (R_\mathrm{e}+H_2)^2,z\geq R_\mathrm{e}|(x,y,z)\in\mathbb{R}^3\}$ and its volume $V_2'$, which can be calculated as
\begin{align}\label{System_Model_8_2}
    V_2'=\frac{\pi}{3}H_2\left(2R_\mathrm{e}+H_2\right)\left(3R_\mathrm{e}+2H_2\right).
\end{align}
Thus, the volume of the visible shell can be calculated as
\begin{align}\label{System_Model_8_3}
    \notag V'=&V_1'-V_2'\\
    \notag=&\frac{\pi}{3}\left(H_2\left(2R_\mathrm{e}+H_2\right)
    \left(3R_\mathrm{e}+2H_2\right)\right.\\
    &\left.-H_1\left(2R_\mathrm{e}+H_1\right)\left(3R_\mathrm{e}+2H_1\right)\right).
\end{align}

\begin{theorem}\label{Theorem_1}
The CDF of the distance $F_{D'}(d)$ from any specific satellite in the visible shell $\Omega'$ to the terrestrial user can be formulated as shown in (\ref{System_Model_9}), when $\sqrt{H_1(2R_\mathrm{e}+H_1)}\leq H_2$, and as shown in (\ref{System_Model_10}) otherwise.

\begin{figure*}[!t]
\begin{align}\label{System_Model_9}
    F_{D'}(d)=\left\{\begin{array}{ll}
        0, & d\in[0,H_1) \\
        \frac{\pi}{4R_\mathrm{e}V'}d^4+\frac{2\pi}{3V'}d^3
        -\frac{\pi H_1\left(2R_\mathrm{e}+H_1\right)}{2R_\mathrm{e}V'}d^2
        +\frac{\pi H_1^3\left(4R_\mathrm{e}+3H_1\right)}{12R_\mathrm{e}V'},
        & d\in[H_1,\sqrt{H_1\left(2R_\mathrm{e}+H_1\right)}) \\
        \frac{2\pi}{3V'}d^3-\frac{\pi H_1^2\left(3R_\mathrm{e}+2H_1\right)}{3V'},
        & d\in[\sqrt{H_1\left(2R_\mathrm{e}+H_1\right)},H_2) \\
        -\frac{\pi}{4R_\mathrm{e}V'}d^4
        +\frac{\pi H_2\left(2R_\mathrm{e}+H_2\right)}{2R_\mathrm{e}V'}d^2
        -\frac{\pi\left(H_2^3\left(4R_\mathrm{e}+3H_2\right)+
        4R_\mathrm{e}H_1^2\left(3R_\mathrm{e}+2H_1\right)\right)}{12R_\mathrm{e}V'},
        & d\in[H_2,\sqrt{H_2\left(2R_\mathrm{e}+H_2\right)}) \\
        1, & d\in[\sqrt{H_2\left(2R_\mathrm{e}+H_2\right)},\infty) \\
        \end{array} \right.
\end{align}
\hrulefill
\end{figure*}

\begin{figure*}[!t]
\begin{align}\label{System_Model_10}
    F_{D'}(d)=\left\{\begin{array}{ll}
        0, & d\in[0,H_1) \\
        \frac{\pi}{4R_\mathrm{e}V'}d^4+\frac{2\pi}{3V'}d^3
        -\frac{\pi H_1\left(2R_\mathrm{e}+H_1\right)}{2R_\mathrm{e}V'}d^2
        +\frac{\pi H_1^3\left(4R_\mathrm{e}+3H_1\right)}{12R_\mathrm{e}V'},
        & d\in[H_1,H_2) \\
        \frac{\pi H_2\left(2R_\mathrm{e}+H_2\right)}{2R_\mathrm{e}V'}d^2
        -\frac{\pi \left(H_2^3\left(4R_\mathrm{e}+3H_2\right)
        -H_1^3\left(4R_\mathrm{e}+3H_1\right)\right)}{12R_\mathrm{e}V'},
        & d\in[H_2,\sqrt{H_1\left(2R_\mathrm{e}+H_1\right)}) \\
        -\frac{\pi}{4R_\mathrm{e}V'}d^4
        +\frac{\pi H_2\left(2R_\mathrm{e}+H_2\right)}{2R_\mathrm{e}V'}d^2
        -\frac{\pi\left(H_2^3\left(4R_\mathrm{e}+3H_2\right)+
        4R_\mathrm{e}H_1^2\left(3R_\mathrm{e}+2H_1\right)\right)}{12R_\mathrm{e}V'},
        & d\in[\sqrt{H_1\left(2R_\mathrm{e}+H_1\right)},
        \sqrt{H_2\left(2R_\mathrm{e}+H_2\right)}) \\
        1, & d\in[\sqrt{H_2\left(2R_\mathrm{e}+H_2\right)},\infty) \\
        \end{array} \right.
\end{align}
\hrulefill
\end{figure*}
\end{theorem}
\begin{IEEEproof}
    See Appendix~\ref{Appendix_A}.
\end{IEEEproof}

\begin{corollary}\label{Corollary_1}
By taking the derivative of the CDF of the distance from any specific satellite in the visible shell $\Omega'$ to the terrestrial user as described in Theorem~\ref{Theorem_1}, we arrive at its PDF formulated as
\begin{align}\label{System_Model_11}
    f_{D'}(d)=\left\{ \begin{array}{ll}
        \frac{\pi}{R_\mathrm{e}V'}d^3
        +\frac{2\pi}{V'}d^2-\frac{\pi \tilde{H}_1^2}{R_\mathrm{e}V'}d,
        & d\in[H_1,\tilde{H}_1) \\
        \frac{2\pi}{V'}d^2,
        & d\in[\tilde{H}_1,H_2) \\
        -\frac{\pi}{R_\mathrm{e}V'}d^3+\frac{\pi \tilde{H}_2^2}{R_\mathrm{e}V'}d,
        & d\in[H_2,\tilde{H}_2) \\
        0, & d\notin[H_1,\tilde{H}_2) \\
        \end{array} \right.,
\end{align}
when $\sqrt{H_1\left(2R_\mathrm{e}+H_1\right)}\leq H_2$ and as
\begin{align}\label{System_Model_12}
    f_{D'}(d)=\left\{ \begin{array}{ll}
        \frac{\pi}{R_\mathrm{e}V'}d^3+\frac{2\pi}{V'}d^2
        -\frac{\pi \tilde{H}_1^2}{R_\mathrm{e}V'}d,
        & d\in[H_1,H_2) \\
        \frac{\pi \tilde{H}_2^2}{R_\mathrm{e}V'}d,
        & d\in[H_2,\tilde{H}_1) \\
        -\frac{\pi}{R_\mathrm{e}V'}d^3+\frac{\pi \tilde{H}_2^2}{R_\mathrm{e}V'}d,
        & d\in[\tilde{H}_1,\tilde{H}_2) \\
        0, & d\notin[H_1,\tilde{H}_2) \\
        \end{array} \right.,
\end{align}
otherwise. In (\ref{System_Model_11}) and (\ref{System_Model_12}), we have $\tilde{H}_1=\sqrt{2R_\mathrm{e}H_1+H_1^2}$ and $\tilde{H}_2=\sqrt{2R_\mathrm{e}H_2+H_2^2}$.
\end{corollary}

Then, we derive the probability of the serving satellite being in the visible shell $\Omega'$ as follows.

\begin{corollary}\label{Corollary_2}
The probability $P_\mathrm{S}$ of the serving satellite being in the visible shelling can be expressed as
\begin{align}\label{System_Model_13}
    P_\mathrm{S}=1-\left(1-P'\right)^{|\mathcal{A}|},
\end{align}
where $P'$ denotes the probability of any specific satellite being in the visible shell $\Omega'$, given by
\begin{align}\label{System_Model_13_1}
    P'=\frac{H_2^2\left(3R_\mathrm{e}+2H_2\right)-H_1^2\left(3R_\mathrm{e}+2H_1\right)}
    {4\left(\left(R_\mathrm{e}+H_2\right)^3-\left(R_\mathrm{e}+H_1\right)^3\right)}.
\end{align}
\end{corollary}
\begin{IEEEproof}
    According to (\ref{System_Model_8_3}), the probability of any specific satellite being in the visible shell $\Omega'$ can be expressed as
    \begin{align}\label{Eq_Corollary_1}
        P'=\frac{V'}{V}.
    \end{align}
    In (\ref{Eq_Corollary_1}), $V$ represents the volume of the shell $\Omega$, given by
    \begin{align}\label{Eq_Corollary_2}
        V=\frac{H_2^2\left(3R_\mathrm{e}+2H_2\right)-H_1^2\left(3R_\mathrm{e}+2H_1\right)}
        {4\left(\left(R_\mathrm{e}+H_2\right)^3-\left(R_\mathrm{e}+H_1\right)^3\right)},
    \end{align}
    and $V'$ is the volume of the visible shell $\Omega'$, as shown in (\ref{System_Model_8_3}). According to (\ref{System_Model_8_3}), (\ref{Eq_Corollary_1}) and (\ref{Eq_Corollary_2}), $P'$ can be expressed as shown in (\ref{System_Model_13_1}).

    Among all $|\mathcal{A}|$ satellites, the serving satellite is the one having the minimal distance from the terrestrial user. Therefore, the probability of the serving satellite located in the visible region can be formulated as shown in (\ref{System_Model_13}).
\end{IEEEproof}

\begin{corollary}\label{Corollary_3}
Given that the serving satellite is located in the visible shell $\Omega'$ and the distance between the serving satellite and the terrestrial user is $D_0=d_0$, the probability $P_\mathrm{I}$ that one of the other satellites is within the visible shell, can be formulated as
\begin{align}\label{System_Model_14}
    P_\mathrm{I}=\frac{\frac{\pi}{3}\left(H_2^2\left(3R_\mathrm{e}+2H_2\right)
    -H_1^2\left(3R_\mathrm{e}+2H_1\right)\right)-V'F_{D'}(d_0)}
    {\frac{4\pi}{3}\left(\left(R_\mathrm{e}+H_2\right)^3-\left(R_\mathrm{e}+H_1\right)^3\right)
    -V'F_{D'}(d_0)}.
\end{align}
\end{corollary}
\begin{IEEEproof}
    Given that the distance between the serving satellite and the terrestrial user is $D_0=d_0$, the probability that one of the interfering satellites is within the visible shell $\Omega'$ is
    \begin{align}\label{Eq_Corollary_3}
        P_\mathrm{I}=\frac{V'-V_{B\cap\Omega'}(d)}{V}=\frac{V'-F_{D'}(d)V'}{V}.
    \end{align}
    Upon substituting (\ref{System_Model_8_3}), (\ref{System_Model_9}), (\ref{System_Model_10}) and (\ref{Eq_Corollary_2}) into (\ref{Eq_Corollary_3}), $P_\mathrm{I}$ can be expressed as seen in (\ref{System_Model_14}).
\end{IEEEproof}

\begin{theorem}\label{Theorem_2}
Given that the distance between the serving satellite and the terrestrial user is $D_0=d_0$, the conditional PDF $f_{D_\mathrm{I}'|D_0}(d|d_0)$ of the distance between any of the interfering satellites located in the visible shell $\Omega'$ and the terrestrial user can be formulated as
\begin{align}\label{System_Model_15}
    f_{D_\mathrm{I}'|D_0}(d|d_0)=\left\{ \begin{array}{ll}
        \frac{\frac{\pi}{R_\mathrm{e}V'}d^3
        +\frac{2\pi}{V'}d^2-\frac{\pi \tilde{H}_1^2}{R_\mathrm{e}V'}d}{1-F_{D'}(d_0)},
        & d\in[H_1,\tilde{H}_1) \\
        \frac{\frac{2\pi}{V'}d^2}{1-F_{D'}(d_0)},
        & d\in[\tilde{H}_1,H_2) \\
        \frac{-\frac{\pi}{R_\mathrm{e}V'}d^3+\frac{\pi \tilde{H}_2^2}{R_\mathrm{e}V'}d}
        {1-F_{D'}(d_0)},
        & d\in[H_2,\tilde{H}_2) \\
        0, & d\notin[H_1,\tilde{H}_2) \\
        \end{array} \right.,
\end{align}
when $\sqrt{H_1(2R_\mathrm{e}+H_1)}\leq H_2$, and
\begin{align}\label{System_Model_16}
    f_{D_\mathrm{I}'|D_0}(d|d_0)=\left\{ \begin{array}{ll}
        \frac{\frac{\pi}{R_\mathrm{e}V'}d^3
        +\frac{2\pi}{V'}d^2-\frac{\pi \tilde{H}_1^2}{R_\mathrm{e}V'}d}{1-F_{D'}(d_0)},
        & d\in[H_1,H_2) \\
        \frac{\frac{\pi \tilde{H}_2^2}{R_\mathrm{e}V'}d}{1-F_{D'}(d_0)},
        & d\in[H_2,\tilde{H}_1) \\
        \frac{-\frac{\pi}{R_\mathrm{e}V'}d^3+\frac{\pi \tilde{H}_2^2}{R_\mathrm{e}V'}d}
        {1-F_{D'}(d_0)},
        & d\in[\tilde{H}_1,\tilde{H}_2) \\
        0, & d\notin[H_1,\tilde{H}_2) \\
        \end{array} \right.,
\end{align}
otherwise.
\end{theorem}
\begin{IEEEproof}
    See Appendix~\ref{Appendix_B}.
\end{IEEEproof}

\section{Hybrid Holographic and Digital Beamforming Design}\label{Beamforming_Design}
In this section, we design the metasurface-based holographic beamformer and the digital beamformer at the terrestrial user for maximizing the throughput of the LEO system considered.

According to (\ref{System_Model_1}), (\ref{System_Model_2}), (\ref{System_Model_3}) and (\ref{System_Model_6}), the baseband channel spanning from the $l$th satellite to the RF chains of the terrestrial user is given by
\begin{align}\label{Beamforming_Design_1}
    \mathbf{h}^{(l)}=\mathbf{A}\mathbf{F}'^{(l)},
\end{align}
where $\mathbf{A}\in\mathbb{C}^{M\times MN}$ is formulated as
\begin{align}\label{Beamforming_Design_2}
    \mathbf{A}=\left[\begin{array}{cccc}
        \mathbf{q}^{(1)\mathrm{T}}\mathbf{B}^{(1)} & \mathbf{0}_{N}^\mathrm{T} & \cdots & \mathbf{0}_{N}^\mathrm{T} \\
        \mathbf{0}_{N}^\mathrm{T} & \mathbf{q}^{(2)\mathrm{T}}\mathbf{B}^{(2)} & \cdots & \mathbf{0}_{N}^\mathrm{T} \\
        \vdots & \vdots & \ddots & \vdots \\
        \mathbf{0}_{N}^\mathrm{T} & \mathbf{0}_{N}^\mathrm{T} & \cdots & \mathbf{q}^{(M)\mathrm{T}}\mathbf{B}^{(M)}
    \end{array}\right].
\end{align}
In (\ref{Beamforming_Design_2}), $\mathbf{B}^{(m)}=\mathbf{Diag}\{\beta_1^{(m)},\beta_2^{(m)},\cdots,\beta_N^{(m)}\}$, and $\mathbf{F}'^{(l)}\in\mathbb{C}^{MN\times 1}$ is given by
\begin{align}\label{Beamforming_Design_3}
    \mathbf{F}'^{(l)}
    =\mathbf{C}\left[\begin{array}{c}
        \mathbf{f}^{(l,1)} \\
        \mathbf{f}^{(l,2)} \\
        \vdots \\
        \mathbf{f}^{(l,M)}
    \end{array}\right]
    =\left[\begin{array}{c}
        \mathbf{f}'^{(l,1)} \\
        \mathbf{f}'^{(l,2)} \\
        \vdots \\
        \mathbf{f}'^{(l,M)}
    \end{array}\right],
\end{align}
where $\mathbf{f}'^{(l,m)}\in\mathbb{C}^{N\times 1}$ represents the link spanning from the $l$th satellite to the $m$th microstrip embodying the mutual coupling. According to (\ref{Beamforming_Design_1}), (\ref{Beamforming_Design_2}) and (\ref{Beamforming_Design_3}), the baseband channel $\mathbf{h}^{(l)}$ can be further expressed as
\begin{align}\label{Beamforming_Design_4}
    \mathbf{h}^{(l)}=\left[\begin{array}{c}
        \mathbf{q}^{(1)\mathrm{T}}\mathbf{B}^{(1)}\mathbf{f}'^{(l,1)} \\
        \mathbf{q}^{(2)\mathrm{T}}\mathbf{B}^{(2)}\mathbf{f}'^{(l,2)} \\
        \vdots \\
        \mathbf{q}^{(M)\mathrm{T}}\mathbf{B}^{(M)}\mathbf{f}'^{(l,M)}
    \end{array}\right].
\end{align}

Therefore, the baseband received signal of the terrestrial user, denoted as $\mathbf{y}\in\mathbb{C}^{M\times1}$, can be formulated as
\begin{align}\label{Beamforming_Design_5}
    \notag\mathbf{y}=&\sqrt{\rho}\mathbf{h}^{(0)}s_0+
    \sum_{\mathbf{p}_l\in\Omega'}\sqrt{\rho}\mathbf{h}^{(l)}s_l+\mathbf{A}\mathbf{C}\mathbf{w}\\
    =&\underbrace{\sqrt{\rho}\mathbf{h}^{(0)}s_0}_{\text{Signal at RF-chains}}+
    \underbrace{\sum_{\mathbf{p}_l\in\Omega'}\sqrt{\rho}\mathbf{h}^{(l)}s_l}
    _{\text{Inter-satellite interference at RF-chains}}+\underbrace{\mathbf{w}'}
    _{\text{Additive noise}},
\end{align}
where $\rho$ is the transmitted power of the satellites, $s_l\in\mathbb{C}^{1\times1}$ is the signal transmitted from the $l$th satellite satisfying $\mathbb{E}[|s_l|^2]=1$, and $\mathbf{w}\in\mathbb{C}^{MN\times1}$ is the additive white Gaussian noise (AWGN) at the reconfigurable metasurface elements satisfying $\mathbf{w}\sim\mathcal{CN}(\mathbf{0}_{MN},\sigma_w^2\mathbf{I}_{MN})$. Furthermore, $\mathbf{w}'$ represents the equivalent additive noise at the RF-chains, given by $\mathbf{w}'=\mathbf{A}\mathbf{C}\mathbf{w}$. Therefore, we have $\mathbf{w}'\sim\mathcal{CN}(\mathbf{0}_M,\sigma_w^2\mathbf{A}
\mathbf{C}\mathbf{C}^\mathrm{H}\mathbf{A}^\mathrm{H})$.

According to (\ref{Beamforming_Design_5}), the signal-to-interference-plus-noise ratio (SINR) is given by
\begin{align}\label{Beamforming_Design_6}
    \gamma_0=\frac{\rho|\mathbf{v}^\mathrm{H}\mathbf{h}^{(0)}|^2}
    {\sum_{\mathbf{p}_l\in\Omega'}\rho|\mathbf{v}^\mathrm{H}\mathbf{h}^{(l)}|^2
    +|\mathbf{v}^\mathrm{H}\mathbf{w}'|^2},
\end{align}
where $\mathbf{v}\in\mathbb{C}^{1\times M}$ is the digital combining vector. Thus, the throughput can be represented as
\begin{align}\label{Beamforming_Design_7}
    \notag R=&P_\mathrm{S}\cdot\log_2\left(1+\gamma_0\right)\\
    =&P_\mathrm{S}\cdot\log_2\left(1+\frac{\rho|\mathbf{v}^\mathrm{H}\mathbf{h}^{(0)}|^2}
    {\sum_{\mathbf{p}_l\in\Omega'}\rho|\mathbf{v}^\mathrm{H}\mathbf{h}^{(l)}|^2
    +|\mathbf{v}^\mathrm{H}\mathbf{w}'|^2}\right).
\end{align}

Here, our aim is to optimize the digital beamformer $\mathbf{v}$ and the reconfigurable metasurface element
coefficient matrix $\mathbf{B}$ in order to maximize the throughput $R$. The corresponding optimization problem can be formulated as
\begin{align}\label{Beamforming_Design_8}
    \mathcal{P}\mathrm{1}:&\max_{\mathbf{v},\mathbf{B}}\
    \log_2\left(1+\frac{\rho|\mathbf{v}^\mathrm{H}\mathbf{h}^{(0)}|^2}
    {\sum_{\mathbf{p}_l\in\Omega'}\rho|\mathbf{v}^\mathrm{H}\mathbf{h}^{(l)}|^2
    +|\mathbf{v}^\mathrm{H}\mathbf{w}'|^2}\right)\\
    \text{s.t.}&\ \beta_n^{(m)}=\frac{\jmath+\mathrm{e}^{\jmath\phi_n^{(m)}}}{2},\\
    &\ \phi_n^{(m)}\in[0,2\pi),\ m=1,2,\cdots,M,\ n=1,2,\cdots,N.
\end{align}

Since $\mathcal{P}\mathrm{1}$ is a non-convex problem, we can decouple it into a pair of sub-problems and optimize them separately. Specifically, the metasurface-based holographic beamformer is designed to maximize the baseband channel gain between the serving satellite and the terrestrial user, while the digital beamformer is optimized based on the MMSE detection method to reduce the inter-satellite interference.

\subsection{Holographic Beamformer}
Maximizing the baseband channel gain in metasurface-based holographic beamformers is crucial for precise beam shaping and for achieving high directional gain. This enhances the signal strength and compensates for the path-loss of satellite-to-ground links, making it ideal for robust communication in dense LEO satellite networks~\cite{deng2023reconfigurable}. To maximize the baseband channel gain between the serving satellite and the terrestrial user, the corresponding problem of optimizing the holographic beamformer can be formulated as
\begin{align}\label{Beamforming_Design_9}
    \mathcal{P}\mathrm{2}:&\max_{\mathbf{B}}\
    \left\|\mathbf{h}^{(0)}\right\|^2\\
    \text{s.t.}&\ \beta_n^{(m)}=\frac{\jmath+\mathrm{e}^{\jmath\phi_n^{(m)}}}{2},\\
    &\ \phi_n^{(m)}\in[0,2\pi),\ m=1,2,\cdots,M,\ n=1,2,\cdots,N.
\end{align}
According to (\ref{Beamforming_Design_4}), the baseband channel gain $\|\mathbf{h}^{(0)}\|^2$ can be further reformulated as
\begin{align}\label{Beamforming_Design_10}
    \notag\left\|\mathbf{h}^{(0)}\right\|^2
    =&\sum_{m=1}^{M}\left|\mathbf{q}^{(m)\mathrm{T}}\mathbf{B}^{(m)}\mathbf{f}'^{(0,m)}\right|^2\\
    =&\sum_{m=1}^{M}\sum_{n=1}^{N}\left|q^{(m)}_n\beta^{(m)}_nf'^{(0,m)}_n\right|^2.
\end{align}
Therefore, the problem $\mathcal{P}\mathrm{2}$ can be recast as
\begin{align}\label{Beamforming_Design_11}
    \mathcal{P}\mathrm{3}:&\max_{\mathbf{B}}\
    \sum_{m=1}^{M}\left|\sum_{n=1}^{N}q^{(m)}_n
    \left(\frac{\jmath+\mathrm{e}^{\jmath\phi_n^{(m)}}}{2}\right)f'^{(0,m)}_n\right|^2\\
    \text{s.t.}&\ \phi_n^{(m)}\in[0,2\pi),\ n=1,2,\cdots,N,
\end{align}
for $m=1,2,\cdots,M$. Finally, the closed-form solution for $\mathcal{P}\mathrm{3}$ can be expressed as
\begin{align}\label{Beamforming_Design_12}
    \notag\phi_n^{(m)}=&\angle\left(\sum_{n=1}^{N}q^{(m)}_nf'^{(0,m)}_n\right)
    +\frac{\pi}{2}\\
    &-\left(\angle q^{(m)}_n+\angle f'^{(0,m)}_n\right),\ n=1,2,\cdots,N.
\end{align}

\subsection{Digital Beamformer}
Based on the holographic beamformer in (\ref{Beamforming_Design_12}), the baseband channel between the serving satellite and the terrestrial user can be formulated as
\begin{align}\label{Beamforming_Design_13}
    \mathbf{h}^{(0)}=\left[\begin{array}{c}
        \frac{1}{2}\left(\left|\sum\limits_{n=1}^{N}q^{(1)}_nf'^{(0,1)}_n\right|
        +\sum\limits_{n=1}^{N}\left|q^{(1)}_n\right|\left|f'^{(0,1)}_n\right|\right) \\
        \frac{1}{2}\left(\left|\sum\limits_{n=1}^{N}q^{(2)}_nf'^{(0,2)}_n\right|
        +\sum\limits_{n=1}^{N}\left|q^{(2)}_n\right|\left|f'^{(0,2)}_n\right|\right) \\
        \vdots \\
        \frac{1}{2}\left(\left|\sum\limits_{n=1}^{N}q^{(M)}_nf'^{(0,M)}_n\right|
        +\sum\limits_{n=1}^{N}\left|q^{(M)}_n\right|\left|f'^{(0,M)}_n\right|\right)
    \end{array}\right].
\end{align}
By contrast, the baseband channel between the interfering satellites and the terrestrial user can be characterized by:
\begin{align}\label{Beamforming_Design_14}
    \mathbf{h}^{(l)}=\left[\begin{array}{c}
        \frac{1}{2}\left(\jmath\sum\limits_{n=1}^{N}q^{(1)}_nf'^{(l,1)}_n
        +\sum\limits_{n=1}^{N}q^{(1)}_nf'^{(l,1)}_n\mathrm{e}^{\jmath\phi_n^{(1)}}\right) \\
        \frac{1}{2}\left(\jmath\sum\limits_{n=1}^{N}q^{(2)}_nf'^{(l,2)}_n
        +\sum\limits_{n=1}^{N}q^{(2)}_nf'^{(l,2)}_n\mathrm{e}^{\jmath\phi_n^{(2)}}\right) \\
        \vdots \\
        \frac{1}{2}\left(\jmath\sum\limits_{n=1}^{N}q^{(M)}_nf'^{(l,M)}_n
        +\sum\limits_{n=1}^{N}q^{(M)}_nf'^{(l,M)}_n\mathrm{e}^{\jmath\phi_n^{(M)}}\right)
    \end{array}\right],
\end{align}
for $l=1,2,\cdots,|\mathcal{A}|-1$, with $\phi_n^{(m)}$ optimized in (\ref{Beamforming_Design_12}).

To design the digital beamformer, we employ the MMSE combining methods based on the required amount of CSI as follows.

\subsubsection{MMSE RC method based on full CSI}
Firstly, we focus our attention on the case of the MMSE RC based on the full CSI, assuming that the terrestrial user can acquire the CSI from both the serving satellite and all the interfering satellites. Therefore, the MMSE RC vector based on the full CSI can be designed as follows:
\begin{align}\label{Beamforming_Design_15}
    \mathbf{v}_\mathrm{f}=\left(\mathbf{h}^{(0)}\mathbf{h}^{(0)\mathrm{H}}
    +\mathbf{U}\right)^{-1}\mathbf{h}^{(0)},
\end{align}
which leads to the attainable throughput of
\begin{align}\label{Beamforming_Design_17}
    \notag R=&P_\mathrm{S}\cdot\log_2
    \left(1+\frac{\left|\mathbf{v}_\mathrm{f}^\mathrm{H}\mathbf{h}^{(0)}\right|^2}
    {\mathbf{v}_\mathrm{f}^\mathrm{H}\mathbf{U}\mathbf{v}_\mathrm{f}}\right)\\
    =&P_\mathrm{S}\cdot\log_2\left(1+\mathbf{h}^{(0)\mathrm{H}}
    \mathbf{U}^{-1}\mathbf{h}^{(0)}\right),
\end{align}
where $\mathbf{U}=\sum\limits_{\mathbf{p}_l\in\Omega'}\mathbf{h}^{(l)}\mathbf{h}^{(l)\mathrm{H}}
+\frac{\sigma_w^2}{\rho}\mathbf{A}\mathbf{C}\mathbf{C}^\mathrm{H}\mathbf{A}^\mathrm{H}$.

\subsubsection{MMSE RC method based on the distribution of satellites}
In practical systems, acquiring the full CSI of all interfering satellites is infeasible. Hence, we propose an MMSE RC method based on the distribution of the satellite constellation by harnessing stochastic geometry. More explicitly, our MMSE RC method utilizes only the average number and spatial distribution of interfering satellites within the visible region $\Omega'$ as statistical information. Notably, it does not rely on the full CSI $\mathbf{h}^{(1)},\mathbf{h}^{(2)},\cdots,\mathbf{h}^{(|\mathcal{A}|-1)}$, or on the precise positional details such as azimuth and elevation angles of individual interfering satellites. By focusing on the distributional characteristics of visible interfering satellites, this approach reduces the complexity of channel estimation. Specifically, the MMSE RC vector based on the statistical information can be formulated as
\begin{align}\label{Beamforming_Design_18}
    \mathbf{v}_\mathrm{s}=\left(\mathbf{h}^{(0)}\mathbf{h}^{(0)\mathrm{H}}
    +\mathbf{R}_\mathrm{I}'+\frac{\sigma_w^2}{\rho}
    \mathbf{A}\mathbf{C}\mathbf{C}^\mathrm{H}\mathbf{A}^\mathrm{H}\right)^{-1}\mathbf{h}^{(0)},
\end{align}
where $\mathbf{R}_\mathrm{I}'$ represents the covariance matrix of the baseband channels between all interfering satellites in the visible shell $\Omega'$ and the terrestrial user. Explicitly, $\mathbf{R}_\mathrm{I}'=\mathbb{E}[\sum_{\mathbf{p}_l\in\Omega'}
\mathbf{h}^{(l)}\mathbf{h}^{(l)\mathrm{H}}]$.

\begin{theorem}\label{Theorem_3}
The covariance matrix of the baseband channels spanning from all interfering satellites in the visible domain $\Omega'$ to the terrestrial user can be expressed as
\begin{align}\label{Beamforming_Design_19}
    \notag\mathbf{R}_\mathrm{I}'
    =&\varsigma\zeta\left(\frac{\lambda}{4\pi}\right)^2\left(|\mathcal{A}|-1\right)
    P_\mathrm{I}\mathcal{L}(d_0)\cdot\\
    &\frac{\mathbf{Q}\left(\mathbf{C}\mathbf{C}^\mathrm{H}
    +\left(\mathbf{C}\mathbf{C}^\mathrm{H}\right)\odot\mathbf{I}_{MN}\right)
    \mathbf{Q}^\mathrm{H}}{4},
\end{align}
where $\mathbf{Q}\in\mathbb{C}^{M\times MN}$ is formulated as
\begin{align}\label{Beamforming_Design_19_1}
    \mathbf{Q}=\left[\begin{array}{cccc}
        \mathbf{q}^{(1)\mathrm{T}} & \mathbf{0}_{N}^\mathrm{T} & \cdots & \mathbf{0}_{N}^\mathrm{T} \\
        \mathbf{0}_{N}^\mathrm{T} & \mathbf{q}^{(2)\mathrm{T}} & \cdots & \mathbf{0}_{N}^\mathrm{T} \\
        \vdots & \vdots & \ddots & \vdots \\
        \mathbf{0}_{N}^\mathrm{T} & \mathbf{0}_{N}^\mathrm{T} & \cdots & \mathbf{q}^{(M)\mathrm{T}}
    \end{array}\right],
\end{align}
and $\mathcal{L}(r_0)$ denotes the average small scale fading of the link spanning from each interfering satellite located in the visible shell $\Omega'$ to the terrestrial user, given the distance $D_0=d_0$ from the serving satellite to the terrestrial terminal. If $\alpha=2$, the value of $\mathcal{L}(d_0)$ is given by
\begin{align}\label{Beamforming_Design_20}
    \mathcal{L}(d_0)=\left\{ \begin{array}{ll}
        \frac{\pi}{2R_\mathrm{e}}d_0^2+2\pi d_0
        -\frac{\pi\tilde{H}_1^2}{R_\mathrm{e}}\ln d_0,
        & d\in[H_1,\tilde{H}_1) \\
        2\pi d_0,
        & d\in[\tilde{H}_1,H_2) \\
        -\frac{\pi}{2R_\mathrm{e}}+\frac{\pi}{R_\mathrm{e}}\tilde{H}_2^2\ln d_0,
        & d\in[H_2,\tilde{H}_2) \\
        \end{array} \right.,
\end{align}
when $\sqrt{H_1(2R_\mathrm{e}+H_1)}\leq H_2$, and
\begin{align}\label{Beamforming_Design_21}
    \mathcal{L}(d_0)=\left\{ \begin{array}{ll}
        \frac{\pi}{2R_\mathrm{e}}d_0^2+2\pi d_0-\frac{\pi\tilde{H}_1^2}{R_\mathrm{e}}\ln d_0,
        & d\in[H_1,H_2) \\
        \frac{\pi}{R_\mathrm{e}}\left(\tilde{H}_2^2-\tilde{H}_1^2\right)\ln d_0,
        & d\in[H_2,\tilde{H}_1) \\
        -\frac{\pi}{2R_\mathrm{e}}+\frac{\pi}{R_\mathrm{e}}\tilde{H}_2^2\ln d_0,
        & d\in[\tilde{H}_1,\tilde{H}_2) \\
        \end{array} \right.,
\end{align}
otherwise. Furthermore, if $\alpha=3$, the value of $\mathcal{L}(d_0)$ is given by
\begin{align}\label{Beamforming_Design_22}
    \mathcal{L}(d_0)=\left\{ \begin{array}{ll}
        \frac{\pi}{R_\mathrm{e}}d_0+2\pi\ln d_0
        +\frac{\pi\tilde{H}_1^2}{R_\mathrm{e}}\frac{1}{d_0},
        & d\in[H_1,\tilde{H}_1) \\
        2\pi\ln d_0,
        & d\in[\tilde{H}_1,H_2) \\
        -\frac{\pi}{R_\mathrm{e}}d_0-\frac{\pi}{R_\mathrm{e}}\tilde{H}_2^2\frac{1}{d_0},
        & d\in[H_2,\tilde{H}_2) \\
        \end{array} \right.,
\end{align}
when $\sqrt{H_1(2R_\mathrm{e}+H_1)}\leq H_2$, and
\begin{align}\label{Beamforming_Design_23}
    \mathcal{L}(d_0)=\left\{ \begin{array}{ll}
        \frac{\pi}{R_\mathrm{e}}d_0+2\pi\ln d_0
        +\frac{\pi\tilde{H}_1^2}{R_\mathrm{e}}\frac{1}{d_0},
        & d\in[H_1,H_2) \\
        -\frac{\pi}{R_\mathrm{e}}\left(\tilde{H}_2^2-\tilde{H}_1^2\right)\frac{1}{d_0},
        & d\in[H_2,\tilde{H}_1) \\
        -\frac{\pi}{R_\mathrm{e}}d_0-\frac{\pi}{R_\mathrm{e}}\tilde{H}_2^2\frac{1}{d_0},
        & d\in[\tilde{H}_1,\tilde{H}_2) \\
        \end{array} \right.,
\end{align}
otherwise. Finally, if $\alpha=4$, the value of $\mathcal{L}(d_0)$ is given by
\begin{align}\label{Beamforming_Design_24}
    \mathcal{L}(d_0)=\left\{ \begin{array}{ll}
        \frac{\pi}{R_\mathrm{e}}\ln d_0-2\pi\frac{1}{d_0}
        +\frac{\pi\tilde{H}_1^2}{2R_\mathrm{e}}\frac{1}{d_0^2},
        & d\in[H_1,\tilde{H}_1) \\
        -2\pi\frac{1}{d_0},
        & d\in[\tilde{H}_1,H_2) \\
        -\frac{\pi}{R_\mathrm{e}}\ln d_0-\frac{\pi}{2R_\mathrm{e}}\tilde{H}_2^2\frac{1}{d_0^2},
        & d\in[H_2,\tilde{H}_2) \\
        \end{array} \right.,
\end{align}
when $\sqrt{H_1(2R_\mathrm{e}+H_1)}\leq H_2$, and
\begin{align}\label{Beamforming_Design_25}
    \mathcal{L}(d_0)=\left\{ \begin{array}{ll}
        \frac{\pi}{R_\mathrm{e}}\ln d_0-2\pi\frac{1}{d_0}
        +\frac{\pi\tilde{H}_1^2}{2R_\mathrm{e}}\frac{1}{d_0^2},
        & d\in[H_1,H_2) \\
        -\frac{\pi}{2R_\mathrm{e}}\left(\tilde{H}_2^2-\tilde{H}_1^2\right)\frac{1}{d_0^2},
        & d\in[H_2,\tilde{H}_1) \\
        -\frac{\pi}{R_\mathrm{e}}\ln d_0-\frac{\pi}{2R_\mathrm{e}}\tilde{H}_2^2\frac{1}{d_0^2},
        & d\in[\tilde{H}_1,\tilde{H}_2) \\
        \end{array} \right.,
\end{align}
otherwise. If $\alpha\neq2,3,4$, the value of $\mathcal{L}(d_0)$ is given by
\begin{align}\label{Beamforming_Design_26}
    \mathcal{L}(d_0)=\left\{ \begin{array}{ll}
        \frac{\pi}{\left(4-\alpha\right)R_\mathrm{e}}d_0^{4-\alpha}
        +\frac{2\pi}{3-\alpha}d_0^{3-\alpha}\\
        \qquad -\frac{\pi\tilde{H}_1^2}{\left(2-\alpha\right)R_\mathrm{e}}d_0^{2-\alpha},
        & d\in[H_1,\tilde{H}_1) \\
        \frac{2\pi}{3-\alpha}d_0^{3-\alpha},
        & d\in[\tilde{H}_1,H_2) \\
        -\frac{\pi}{\left(4-\alpha\right)R_\mathrm{e}}d_0^{4-\alpha}\\
        \qquad +\frac{\pi\tilde{H}_2^2}{\left(2-\alpha\right)R_\mathrm{e}}d_0^{2-\alpha},
        & d\in[H_2,\tilde{H}_2) \\
        \end{array} \right.,
\end{align}
when $\sqrt{H_1(2R_\mathrm{e}+H_1)}\leq H_2$, and
\begin{align}\label{Beamforming_Design_27}
    \mathcal{L}(d_0)=\left\{ \begin{array}{ll}
        \frac{\pi}{\left(4-\alpha\right)R_\mathrm{e}}d_0^{4-\alpha}
        +\frac{2\pi}{3-\alpha}d_0^{3-\alpha}\\
        \qquad -\frac{\pi\tilde{H}_1^2}{\left(2-\alpha\right)R_\mathrm{e}}d_0^{2-\alpha},
        & d\in[H_1,H_2) \\
        \frac{\pi}{\left(2-\alpha\right)R_\mathrm{e}}
        \left(\tilde{H}_2^2-\tilde{H}_1^2\right)d_0^{2-\alpha},
        & d\in[H_2,\tilde{H}_1) \\
        -\frac{\pi}{\left(4-\alpha\right)R_\mathrm{e}}d_0^{4-\alpha}\\
        \qquad +\frac{\pi\tilde{H}_2^2}{\left(2-\alpha\right)R_\mathrm{e}}d_0^{2-\alpha},
        & d\in[\tilde{H}_1,\tilde{H}_2) \\
        \end{array} \right.,
\end{align}
otherwise.
\end{theorem}
\begin{IEEEproof}
    See Appendix~\ref{Appendix_C}.
\end{IEEEproof}

Given the satellite distribution, the throughput of the hybrid beamforming method can be formulated as in (\ref{Beamforming_Design_28}).
\begin{figure*}[!t]
\begin{align}\label{Beamforming_Design_28}
    R=P_\mathrm{S}\cdot\log_2
    \left(1+\frac{\left|\mathbf{v}_\mathrm{s}^\mathrm{H}\mathbf{h}^{(0)}\right|^2}
    {\mathbf{v}_\mathrm{s}^\mathrm{H}\mathbf{U}\mathbf{v}_\mathrm{s}}\right)
    =P_\mathrm{S}\cdot\log_2
    \left(1+\frac{\mathbf{h}^{(0)\mathrm{H}}\left(\mathbf{R}_\mathrm{I}'+\frac{\sigma_w^2}{\rho}
    \mathbf{A}\mathbf{C}\mathbf{C}^\mathrm{H}\mathbf{A}^\mathrm{H}\right)^{-1}\mathbf{h}^{(0)}}
    {\mathbf{h}^{(0)\mathrm{H}}\left(\mathbf{R}_\mathrm{I}'+\frac{\sigma_w^2}{\rho}
    \mathbf{A}\mathbf{C}\mathbf{C}^\mathrm{H}\mathbf{A}^\mathrm{H}\right)^{-1}
    \mathbf{U}\left(\mathbf{R}_\mathrm{I}'+\frac{\sigma_w^2}{\rho}
    \mathbf{A}\mathbf{C}\mathbf{C}^\mathrm{H}\mathbf{A}^\mathrm{H}\right)^{-1}
    \mathbf{h}^{(0)}}\right).
\end{align}
\hrulefill
\end{figure*}

\subsection{Computational Complexity of the RC Methods}
The computational complexity of the MMSE RC methods can be quantified in terms of the calculation of
$\mathbf{v}_\mathrm{f}^\mathrm{H}\mathbf{y}$ or $\mathbf{v}_\mathrm{s}^\mathrm{H}\mathbf{y}$ for every received signal $\mathbf{y}$ at the terrestrial user. The complexity of additions and subtractions is neglected, since it is considerably lower. Hence, we quantify the complexity by counting the number of floating-point multiplication and division operations required for the calculation of $\mathbf{v}_\mathrm{f}^\mathrm{H}\mathbf{y}$ or $\mathbf{v}_\mathrm{s}^\mathrm{H}\mathbf{y}$.

\begin{table*}[!t]
\footnotesize
\begin{center}
\caption{Complexity of the MMSE RC methods based on various levels of CSI.}\label{Table_calculation_complexity}
\begin{tabular}{|c|c|c|c|c|c|c|}
\hline
\multirow{2}{*}{\textbf{Schemes}} & \multicolumn{2}{|c|}{\textbf{Calculation of combining vectors}} & \multicolumn{2}{|c|}{\textbf{Information recovery}} & \multicolumn{2}{|c|}{\textbf{Total calculation complexity}} \\
\cline{2-7}
 & Multiplication & Division & Multiplication & Division & Multiplication & Division \\
\hline
MMSE, full CSI & $\left(\left(|\mathrm{A}|-1\right)P_\mathrm{I}+2\right)M^2$ & $M$ & $M$ & 0 & $\left(\left(|\mathrm{A}|-1\right)P_\mathrm{I}+2\right)M^2+\tau M$ & $M$ \\
\hdashline
MMSE, satellite distribution & $2M^2$ & $M$ & $M$ & 0 & $2M^2+\tau M$ & $M$ \\
\hline
\end{tabular}
\end{center}
\end{table*}

The complexity of the MMSE RC is presented in Table~\ref{Table_calculation_complexity}, where $\tau$ denotes the number of information symbols within each coherence interval. Specifically, in the MMSE RC method based on the full CSI, the computational complexity includes the calculation of the combining vector $\mathbf{v}_\mathrm{f}$ in (\ref{Beamforming_Design_15}) and that of the information recovery $\mathbf{v}_\mathrm{f}^\mathrm{H}\mathbf{y}$. Specifically, in terms of $\mathbf{v}_\mathrm{f}$, the calculation of $\mathbf{h}^{(0)}\mathbf{h}^{(0)\mathrm{H}}+
\sum_{\mathbf{p}_l\in\Omega'}\mathbf{h}^{(l)}\mathbf{h}^{(l)\mathrm{H}}$ requires an average number of $[(|\mathrm{A}|-1)P_\mathrm{I}+1]M^2$ floating-point multiplications, while the complexity of calculating $\frac{\sigma_w^2}{\rho}\mathbf{A}\mathbf{C}\mathbf{C}^\mathrm{H}\mathbf{A}^\mathrm{H}$ can be ignored, since it remains unchanged in each statistical block. Furthermore, the inverse of the matrix $\mathbf{h}^{(0)}\mathbf{h}^{(0)\mathrm{H}}+\sum_{\mathbf{p}_l\in\Omega'}
\mathbf{h}^{(l)}\mathbf{h}^{(l)\mathrm{H}}+\frac{\sigma_w^2}{\rho}
\mathbf{A}\mathbf{C}\mathbf{C}^\mathrm{H}\mathbf{A}^\mathrm{H}$ and the multiplication with the vector $\mathbf{h}^{(0)}$ require $M^2$ number of floating-point multiplications and $M$ floating-point divisions by utilizing the $\mathrm{LDL}^\mathrm{H}$ decomposition~\cite{bjornson2017massive}. In terms of the information recovery of $\mathbf{v}_\mathrm{f}^\mathrm{H}\mathbf{y}$, we require $M$ floating-point multiplications. By contrast, in the MMSE combining method based on the statistical CSI, the computational complexity includes the calculation of the combining vector $\mathbf{v}_\mathrm{s}$ in (\ref{Beamforming_Design_18}) and that of the information recovery $\mathbf{v}_\mathrm{s}^\mathrm{H}\mathbf{y}$. Specifically, in terms of $\mathbf{v}_\mathrm{s}$, the calculation of $\mathbf{h}^{(0)}\mathbf{h}^{(0)\mathrm{H}}$ requires the average number of $M^2$ floating-point multiplications, while the computational complexity of calculating $\mathbf{R}_\mathrm{I}'$ and $\frac{\sigma_w^2}{\rho}\mathbf{A}\mathbf{C}\mathbf{C}^\mathrm{H}\mathbf{A}^\mathrm{H}$ can be readily ignored, since they remain unchanged in each statistical block. Furthermore, the inverse of the matrix $\mathbf{h}^{(0)}\mathbf{h}^{(0)\mathrm{H}}+\mathbf{R}_\mathrm{I}'
+\frac{\sigma_w^2}{\rho}\mathbf{A}\mathbf{C}\mathbf{C}^\mathrm{H}\mathbf{A}^\mathrm{H}$ and its multiplication with the vector $\mathbf{h}^{(0)}$ require $M^2$ number of floating-point multiplications and $M$ floating-point divisions by utilizing the $\mathrm{LDL}^\mathrm{H}$ decomposition. In terms of the information recovery of $\mathbf{v}_\mathrm{s}^\mathrm{H}\mathbf{y}$, we require $M$ floating-point multiplications.

In summary, the MMSE RC method based on full CSI requires a higher computational complexity due to the need for complete CSI acquired from both the serving and interfering satellites, which includes calculating the full covariance matrix. By contrast, the MMSE RC method based on the satellite distribution significantly reduces complexity by relying on the statistical characteristics of the satellite constellation, thus avoiding the overhead associated with acquiring full CSI. This approach has lower computational demand, making it more feasible for dense LEO satellite deployments.

\section{Simulation Results}\label{Simulation_Results}
In this section, we numerically evaluate the performance of proposed beamforming methods for the holographic metasurface-based multi-altitude LEO satellite communications. We assume that the holographic metasurface elements are compactly packed, i.e. the physical size of each microstrip is $N\delta_N$. In all simulations, mutual coupling is modeled based on (\ref{System_Model_6}), providing a realistic assessment of the element interactions under dense element configurations. Each dipole is arranged parallel to the holographic surface, ensuring uniform orientation across the planar array. This configuration maximizes surface efficiency and simplifies mutual coupling modeling by allowing a consistent application of the coupling model across the elements. The simulation parameters are similar to those in~\cite{an2024stacked_earlyaccess}, \cite{deng2022reconfigurable_twc}, which are given in Table~\ref{Table_Simulation}, unless specified otherwise.

\begin{table}[!t]
\footnotesize
\begin{center}
\caption{Simulation Parameters.}\label{Table_Simulation}
\begin{tabular}{|c|c|}
\hline
    \multicolumn{1}{|p{4.8cm}|}{\raggedright \textbf{Parameters}} & \multicolumn{1}{|p{3.1cm}|}{\raggedleft \textbf{Values}} \\
\hline
    \multicolumn{1}{|p{4.8cm}|}{\raggedright Carrier frequency} & \multicolumn{1}{|p{3.1cm}|}{\raggedleft $f_c=28$ GHz}\\
\hdashline
    \multicolumn{1}{|p{4.8cm}|}{\raggedright Bandwidth} & \multicolumn{1}{|p{3.1cm}|}{\raggedleft $B_w=1$ MHz}\\
\hdashline
    \multicolumn{1}{|p{4.8cm}|}{\raggedright Antenna gain} & \multicolumn{1}{|p{3.1cm}|}{\raggedleft $\zeta=50$ dBi}\\
\hdashline
    \multicolumn{1}{|p{4.8cm}|}{\raggedright Number of microstrips} & \multicolumn{1}{|p{3.1cm}|}{\raggedleft $M=4$} \\
\hdashline
    \multicolumn{1}{|p{4.8cm}|}{\raggedright Number of elements in each microstrip} & \multicolumn{1}{|p{3.1cm}|}{\raggedleft $N=8$} \\
\hdashline
    \multicolumn{1}{|p{4.8cm}|}{\raggedright Earth radius} & \multicolumn{1}{|p{3.1cm}|}{\raggedleft $R_\mathrm{e}=6371$ km} \\
\hdashline
    \multicolumn{1}{|p{4.8cm}|}{\raggedright Microstrip spacing} & \multicolumn{1}{|p{3.1cm}|}{\raggedleft $\delta_M=\frac{\lambda}{2}$} \\
\hdashline
    \multicolumn{1}{|p{4.8cm}|}{\raggedright Holographic metasurface element spacing} & \multicolumn{1}{|p{3.1cm}|}{\raggedleft $\delta_N=\frac{\lambda}{4}$} \\
\hdashline
    \multicolumn{1}{|p{4.8cm}|}{\raggedright Dipole length} & \multicolumn{1}{|p{3.1cm}|}{\raggedleft $\delta_0=\delta_N$} \\
\hdashline
    \multicolumn{1}{|p{4.8cm}|}{\raggedright Minimum altitude of LEO satellites} & \multicolumn{1}{|p{3.1cm}|}{\raggedleft $H_1=160$ km} \\
\hdashline
    \multicolumn{1}{|p{4.8cm}|}{\raggedright Maximum altitude of LEO satellites} & \multicolumn{1}{|p{3.1cm}|}{\raggedleft $H_2=2000$ km} \\
\hdashline
    \multicolumn{1}{|p{4.8cm}|}{\raggedright Number of satellites} & \multicolumn{1}{|p{3.1cm}|}{\raggedleft $|\mathcal{A}|=720$} \\
\hdashline
    \multicolumn{1}{|p{4.8cm}|}{\raggedright Satellite transmit power} & \multicolumn{1}{|p{3.1cm}|}{\raggedleft $\rho=60$ dBW} \\
\hdashline
    \multicolumn{1}{|p{4.8cm}|}{\raggedright Path loss exponent} & \multicolumn{1}{|p{3.1cm}|}{\raggedleft $\alpha=2$} \\
\hdashline
    \multicolumn{1}{|p{4.8cm}|}{\raggedright Average rain attenuation} & \multicolumn{1}{|p{3.1cm}|}{\raggedleft $\varsigma=-4.324$ dB} \\
\hdashline
    \multicolumn{1}{|p{4.8cm}|}{\raggedright Noise power} & \multicolumn{1}{|p{3.1cm}|}{\raggedleft $\sigma_w^2=-104$ dBm} \\
\hdashline
    \multicolumn{1}{|p{4.8cm}|}{\raggedright Shadowed-Rician fading coefficients} & \multicolumn{1}{|p{3.1cm}|}{\raggedleft $b=0.3$, $\upsilon=3$, $\omega=0.4$} \\
\hline
\end{tabular}
\end{center}
\end{table}

\begin{figure*}[!t]
    \centering
    \subfloat[Number of microstrips $M=2$.]
    {\begin{minipage}{0.33\linewidth}
        \centering
        \includegraphics[width=2.35in]{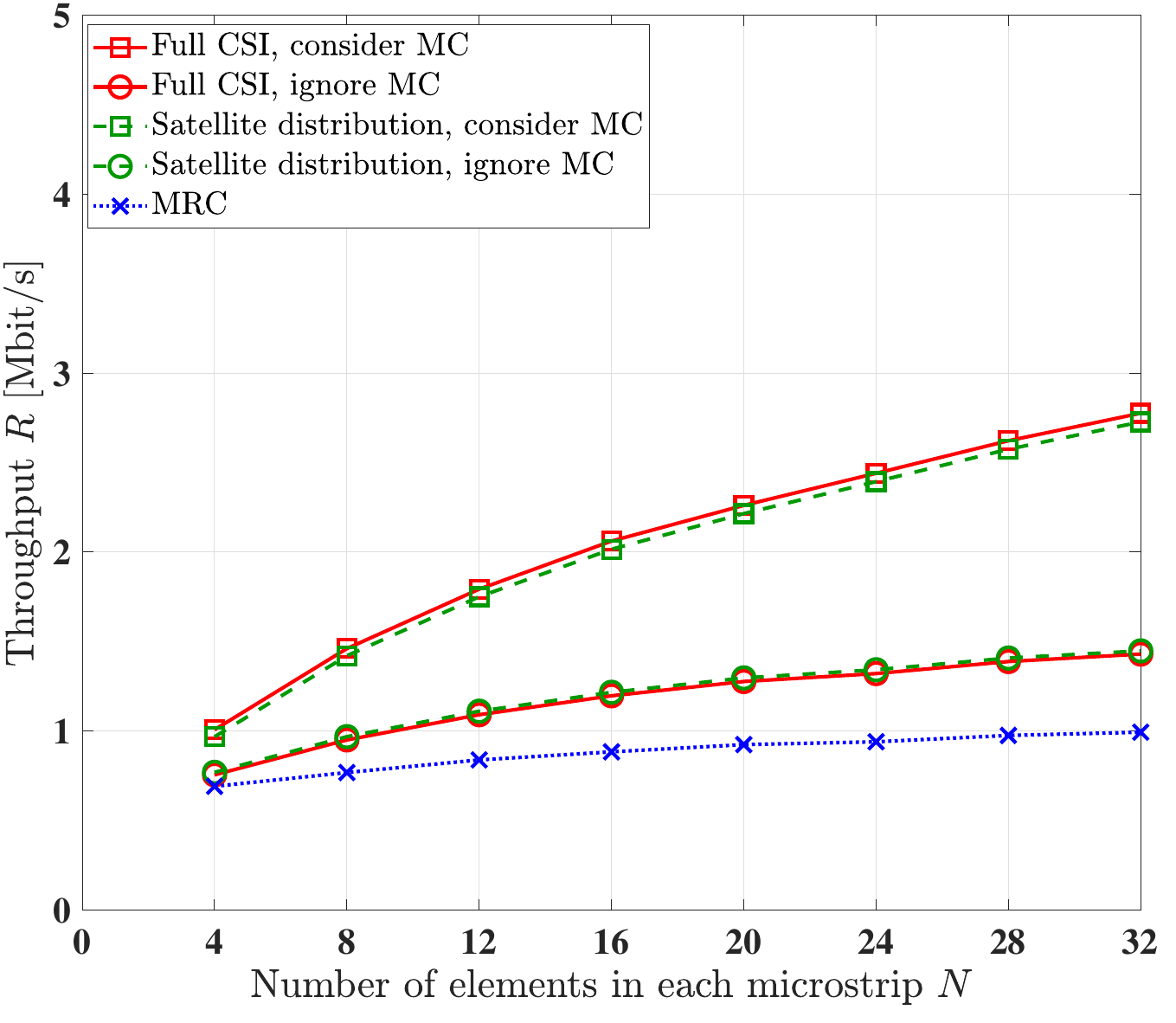}
    \end{minipage}}
    \subfloat[Number of microstrips $M=4$.]
    {\begin{minipage}{0.33\linewidth}
        \centering
        \includegraphics[width=2.35in]{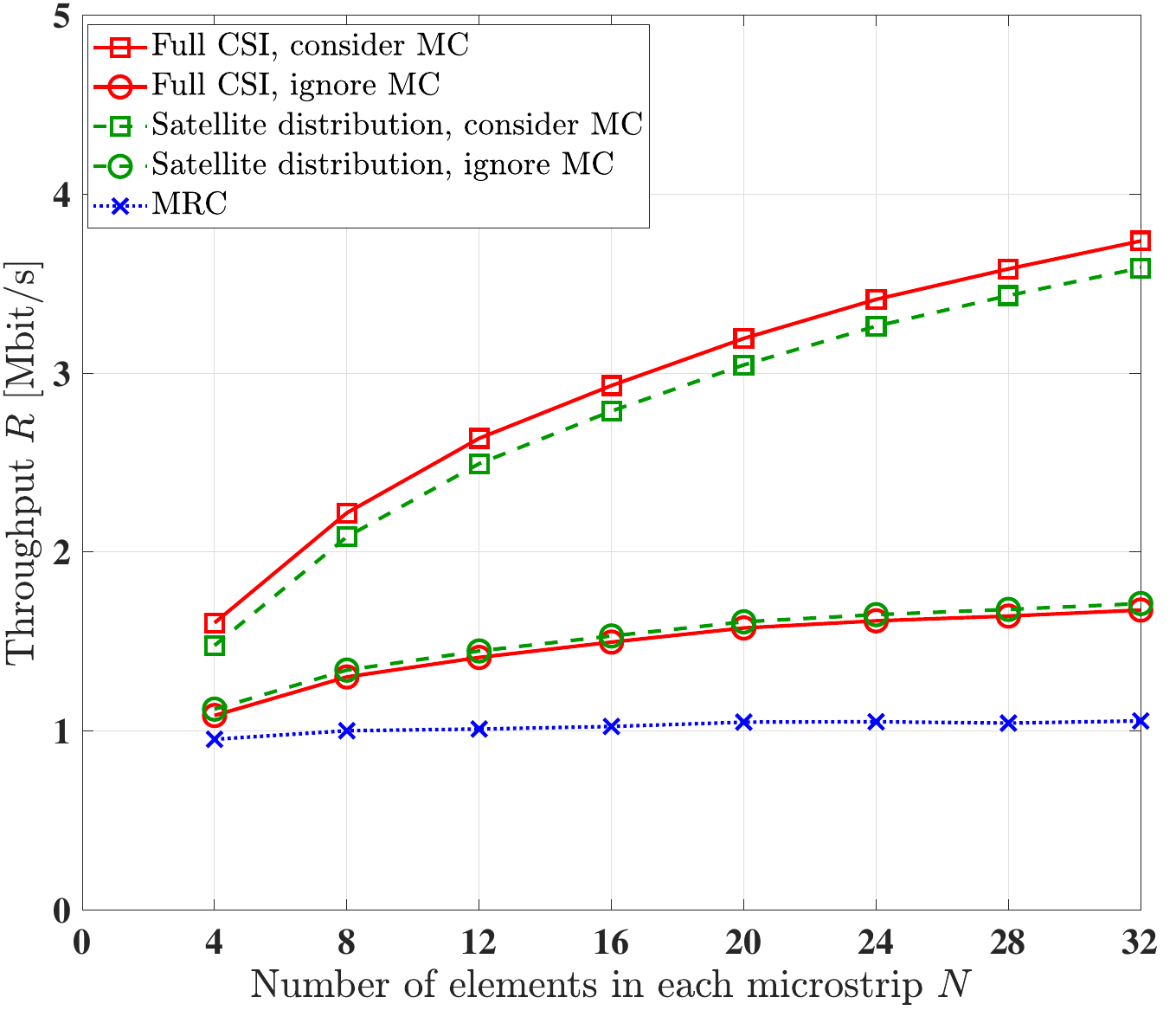}
    \end{minipage}}
    \subfloat[Number of microstrips $M=6$.]
    {\begin{minipage}{0.33\linewidth}
        \centering
        \includegraphics[width=2.35in]{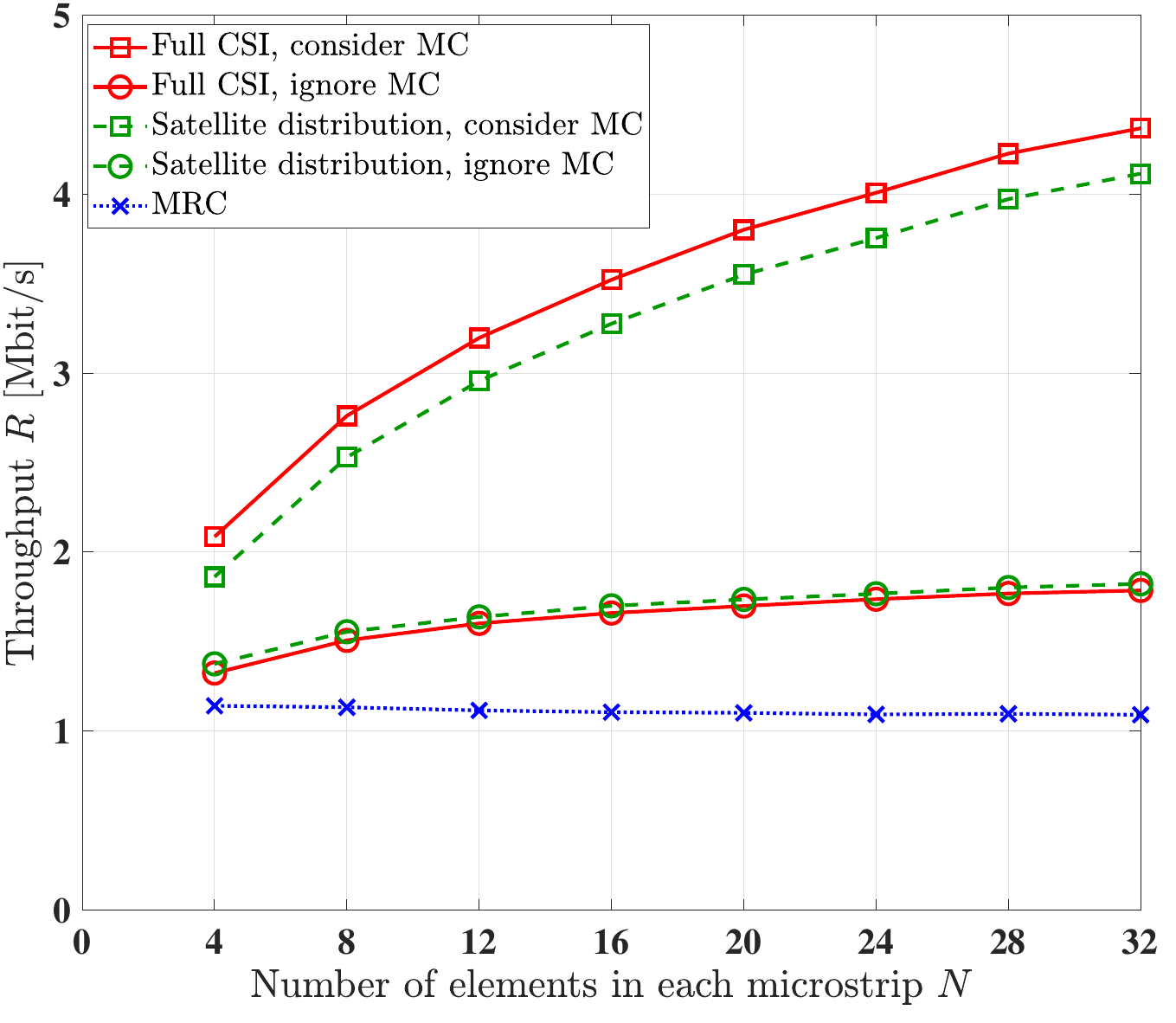}
    \end{minipage}}
    \caption{Throughput $R$ versus the number of holographic metasurface elements in each microstrip $N$, with the fixed holographic metasurface element spacing $\delta_N=\frac{\lambda}{4}$ in each microstrip, i.e. the physical dimension of each microstrip being $\frac{\lambda}{4}N$.}\label{Simu_figure_123}
\end{figure*}

\begin{figure*}[!t]
    \centering
    \subfloat[Number of microstrips $M=2$.]
    {\begin{minipage}{0.33\linewidth}
        \centering
        \includegraphics[width=2.35in]{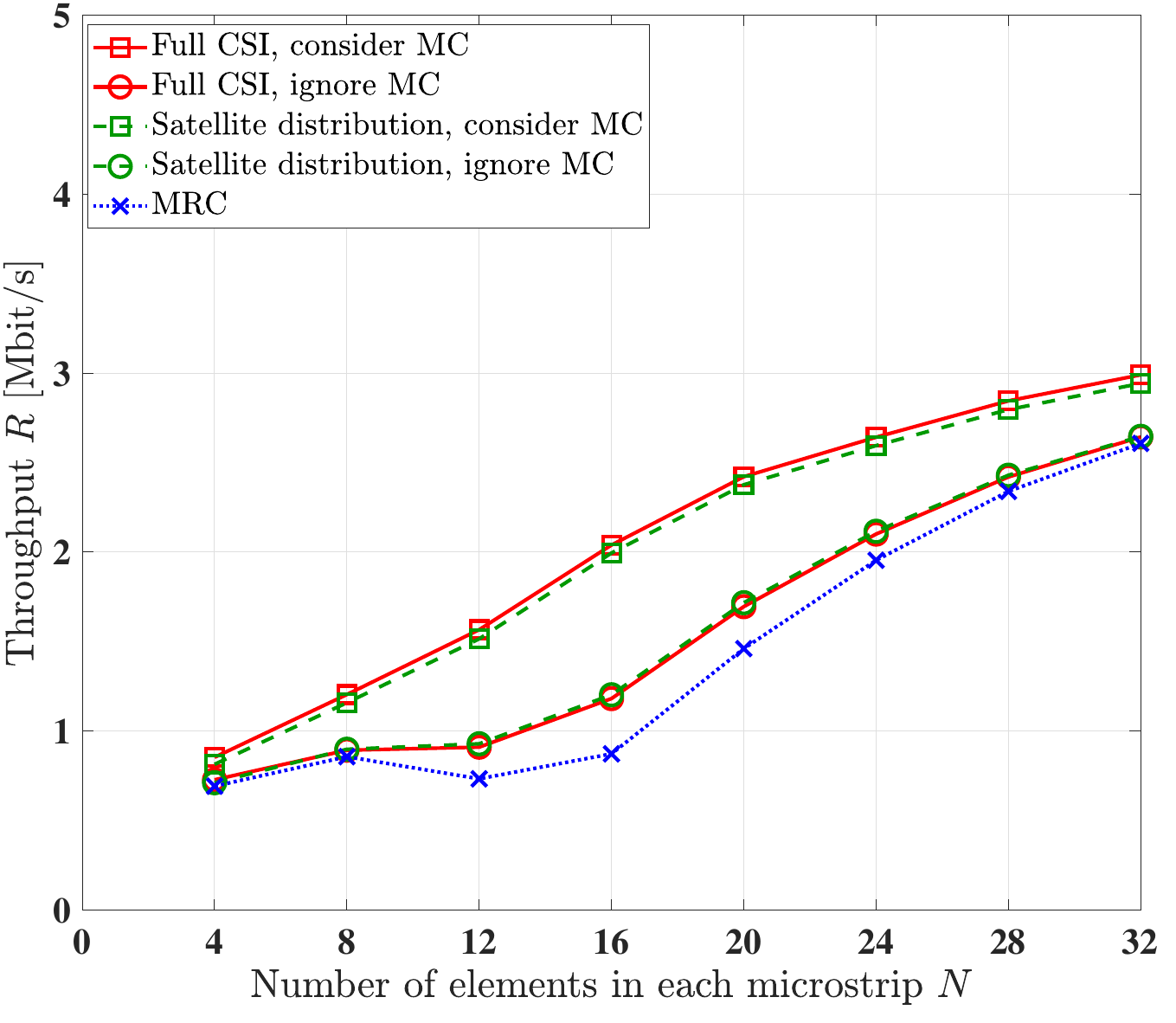}
    \end{minipage}}
    \subfloat[Number of microstrips $M=4$.]
    {\begin{minipage}{0.33\linewidth}
        \centering
        \includegraphics[width=2.35in]{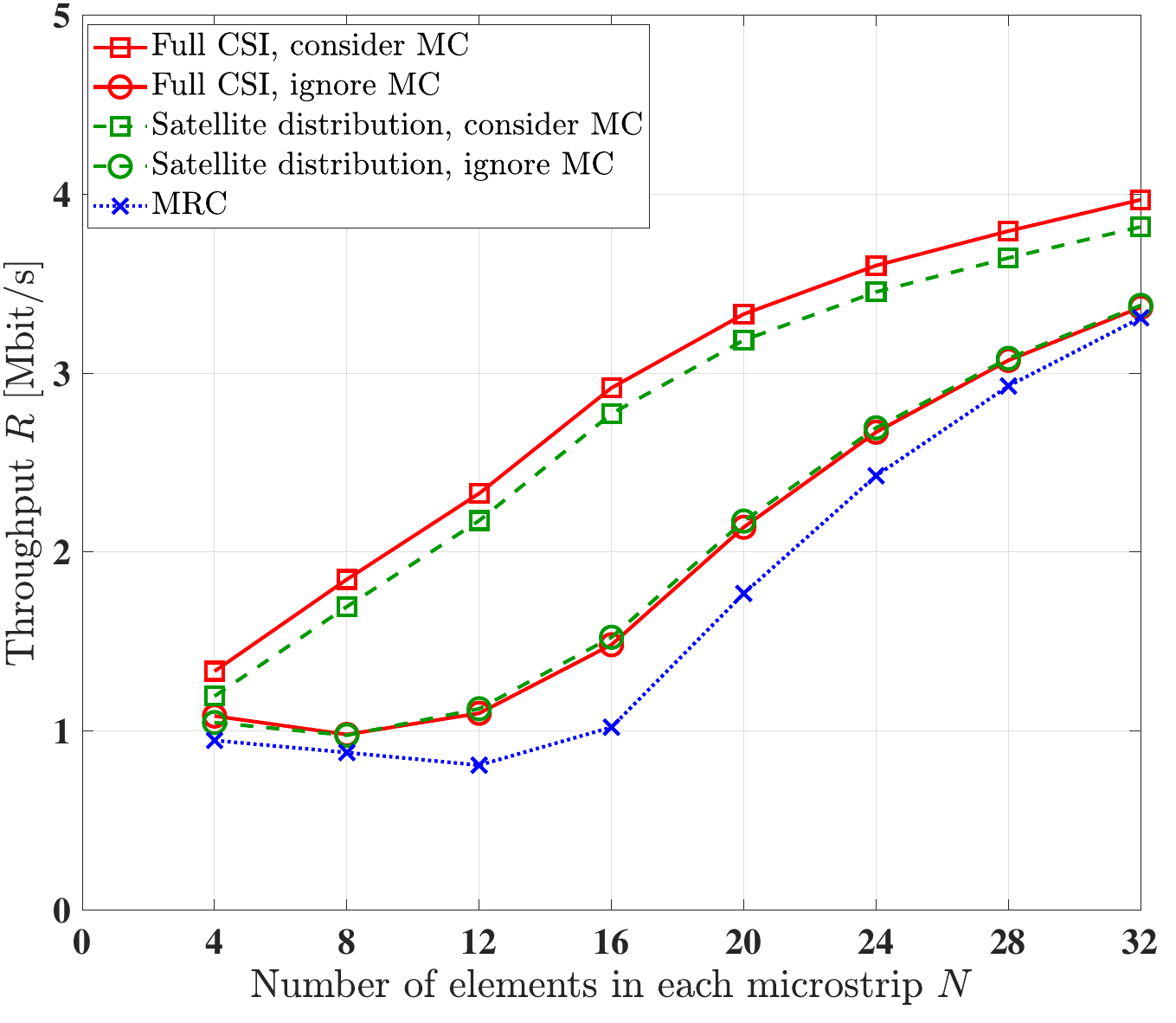}
    \end{minipage}}
    \subfloat[Number of microstrips $M=6$.]
    {\begin{minipage}{0.33\linewidth}
        \centering
        \includegraphics[width=2.35in]{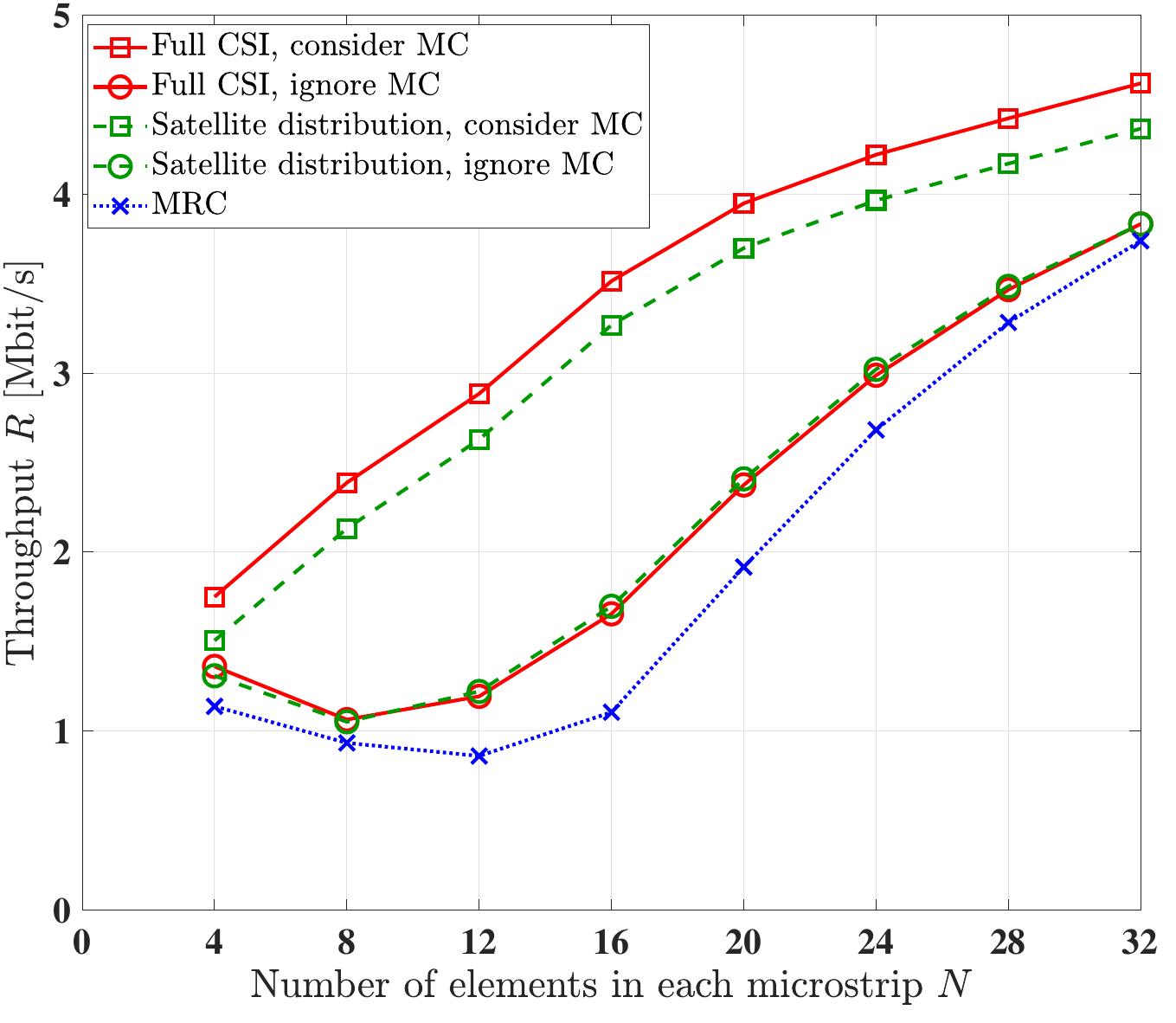}
    \end{minipage}}
    \caption{Throughput $R$ versus the number of holographic metasurface elements in each microstrip $N$, with the fixed physical dimension of $4\lambda$ for each microstrip, i.e. for the holographic metasurface element spacing of $\delta_N=\frac{4\lambda}{N}$.}\label{Simu_figure_456}
\end{figure*}

Fig.~\ref{Simu_figure_123} compares the throughput $R$ versus the number $N$ of holographic metasurface elements in each microstrip. The holographic metasurface element spacing is fixed to $\delta_N=\frac{\lambda}{4}$ in each microstrip, which means that the physical dimension of each microstrip is $\frac{\lambda}{4}N$. The legend `consider MC' indicates considering the effect of mutual coupling in the beamforming design, while `ignore MC' means that the mutual coupling is present among the holographic metasurface elements but not accounted for in the beamforming designs. Observe in Fig.~\ref{Simu_figure_123} that as expected, the throughput can be significantly improved upon considering the effect of mutual coupling in the beamforming design. Furthermore, the MMSE RC method based on the satellite distribution can achieve almost the same throughput as that based on the full CSI, even though it has a lower overhead and a lower computational complexity. Compared to the MRC method, where the RC vector is designed as $\mathbf{v}=\mathbf{h}^{(0)}$, the MMSE RC method exhibits higher throughput since it can effectively mitigate the inter-satellite interference.

Fig.~\ref{Simu_figure_456} portrays the throughput $R$ versus the number of holographic metasurface elements $N$ in each microstrip, with a fixed physical dimension of $4\lambda$ per microstrip. This setup implies that the holographic metasurface element spacing is $\delta_N=\frac{4\lambda}{N}$. Fig.~\ref{Simu_figure_456} demonstrates that increasing the number of holographic metasurface elements in each microstrip enhances the throughput. Moreover, employing more RF chains can further boost the throughput at the expense of increased hardware and energy costs. Notably, the throughput may degrade as the number of elements in each microstrip increases when the effect of mutual coupling is ignored in the beamforming design. This degradation occurs due to the intensified mutual coupling caused by the reduced spacing between metasurface elements.

\begin{figure}[!t]
    \centering
    \subfloat[MMSE method based on the full CSI.]
    {\begin{minipage}{1\linewidth}
        \centering
        \includegraphics[width=2.35in]{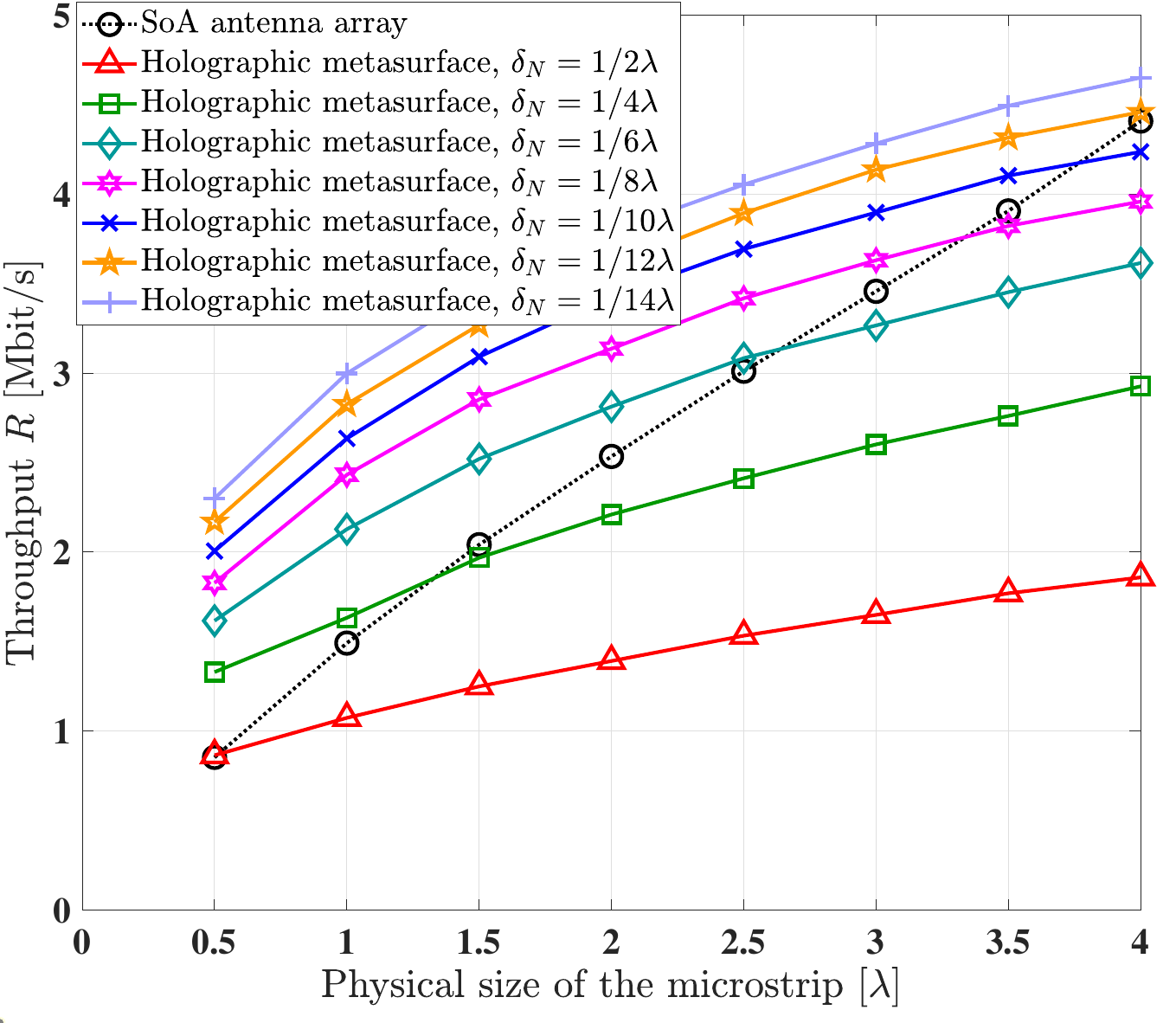}
    \end{minipage}}\\
    \subfloat[MMSE method based on the statistical CSI.]
    {\begin{minipage}{1\linewidth}
        \centering
        \includegraphics[width=2.35in]{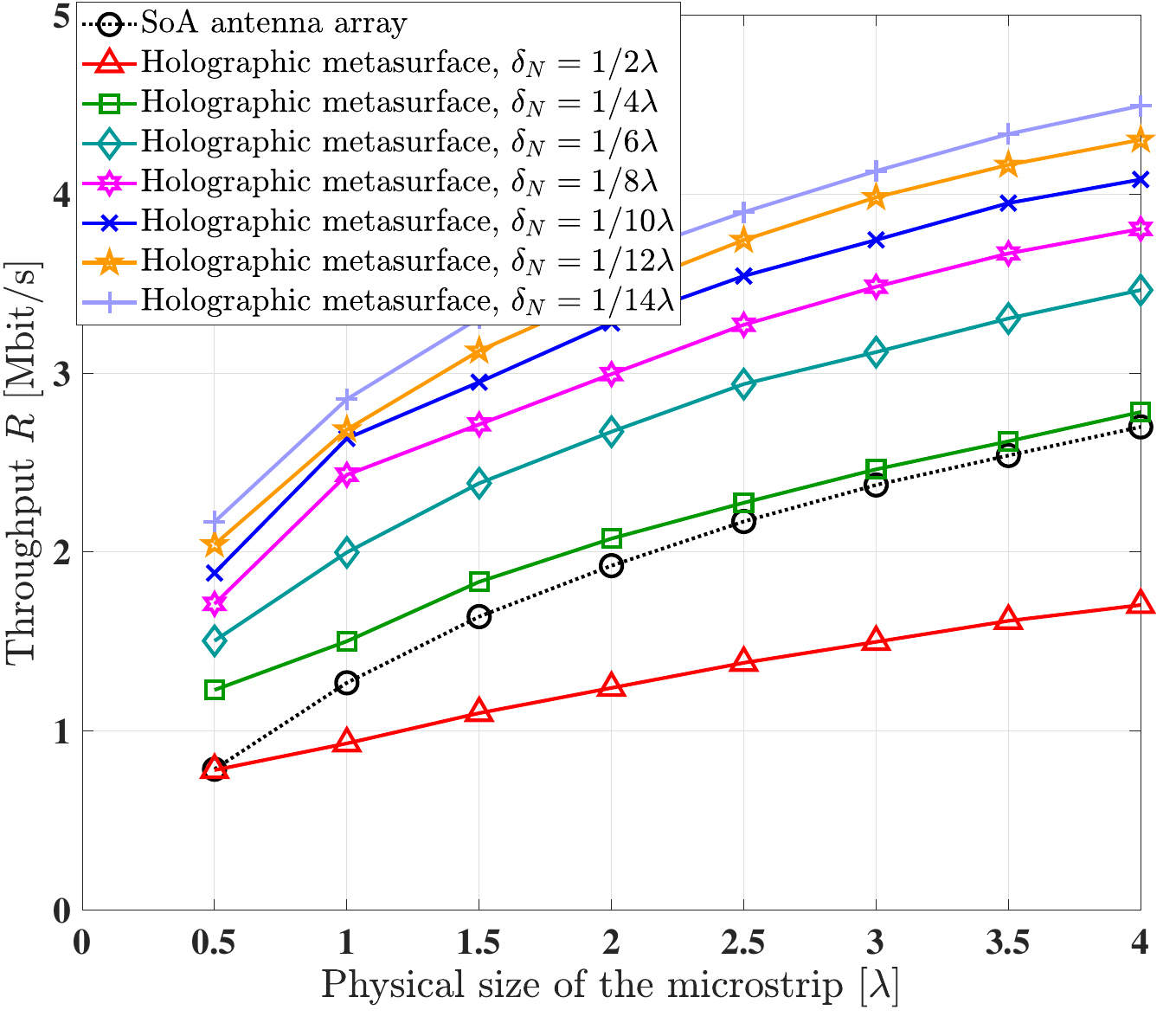}
    \end{minipage}}
    \caption{The throughput $R$ versus the physical size of the microstrip.}\label{Simu_figure_78}
\end{figure}

To provide further insights, Fig.~\ref{Simu_figure_78} illustrates the throughput $R$ as a function of the physical size of the microstrip, considering different holographic metasurface element spacings. Observe that the throughput can be improved by decreasing the holographic metasurface element spacing, which allows for more elements to be employed within the fixed physical size of the microstrip. Additionally, the throughput of the holographic metasurface is compared to that of the SoA antenna array architecture. In the latter, the antenna spacing is $\frac{\lambda}{2}$ and full-digital beamforming is employed. The results indicate that the holographic metasurface outperforms the SoA antenna array upon decreasing the holographic metasurface element spacing.

\begin{figure}[!t]
    \centering
    \includegraphics[width=2.35in]{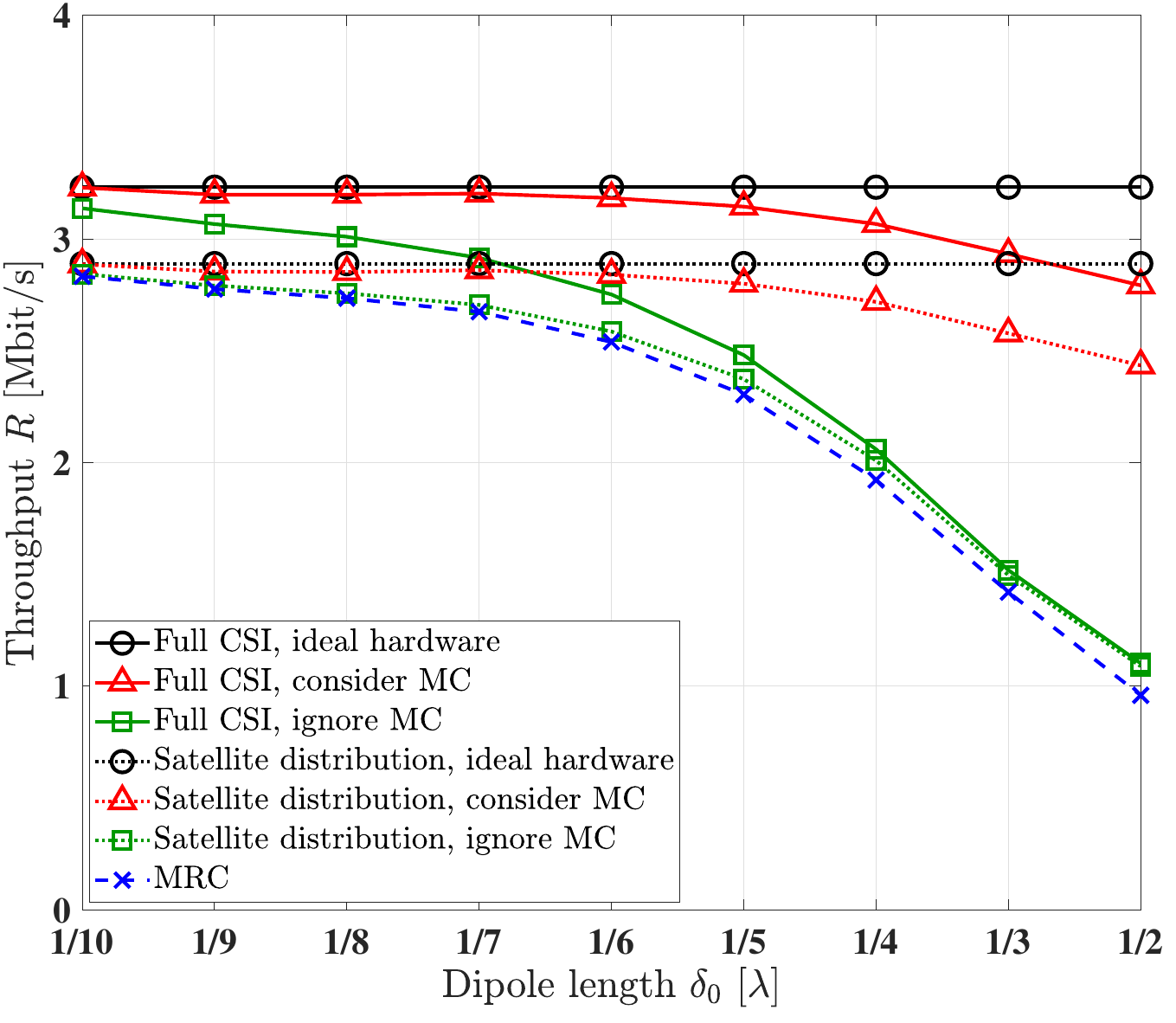}
    \caption{The throughput $R$ versus the dipole length $\delta_0$ in various beamforming methods.}\label{Simu_figure_9}
\end{figure}

Fig.~\ref{Simu_figure_9} compares the throughput $R$ versus the dipole length $\delta_0$ across various detection methods. The legend `ideal hardware' refers to the absence of mutual coupling among holographic metasurface elements, i.e., when the mutual coupling matrix obeys $\mathbf{C}=\mathbf{I}_{MN}$. Observe that the throughput of the holographic metasurface relying on non-ideal hardware approaches that of the ideal hardware when the dipole length is small, as the side effects of mutual coupling on the system performance are reduced. By contrast, when the dipole length is long, mutual coupling significantly reduces the throughput. However, incorporating mutual coupling into the beamforming design can effectively compensate for the reduced throughput.

\begin{figure}[!t]
    \centering
    \includegraphics[width=2.35in]{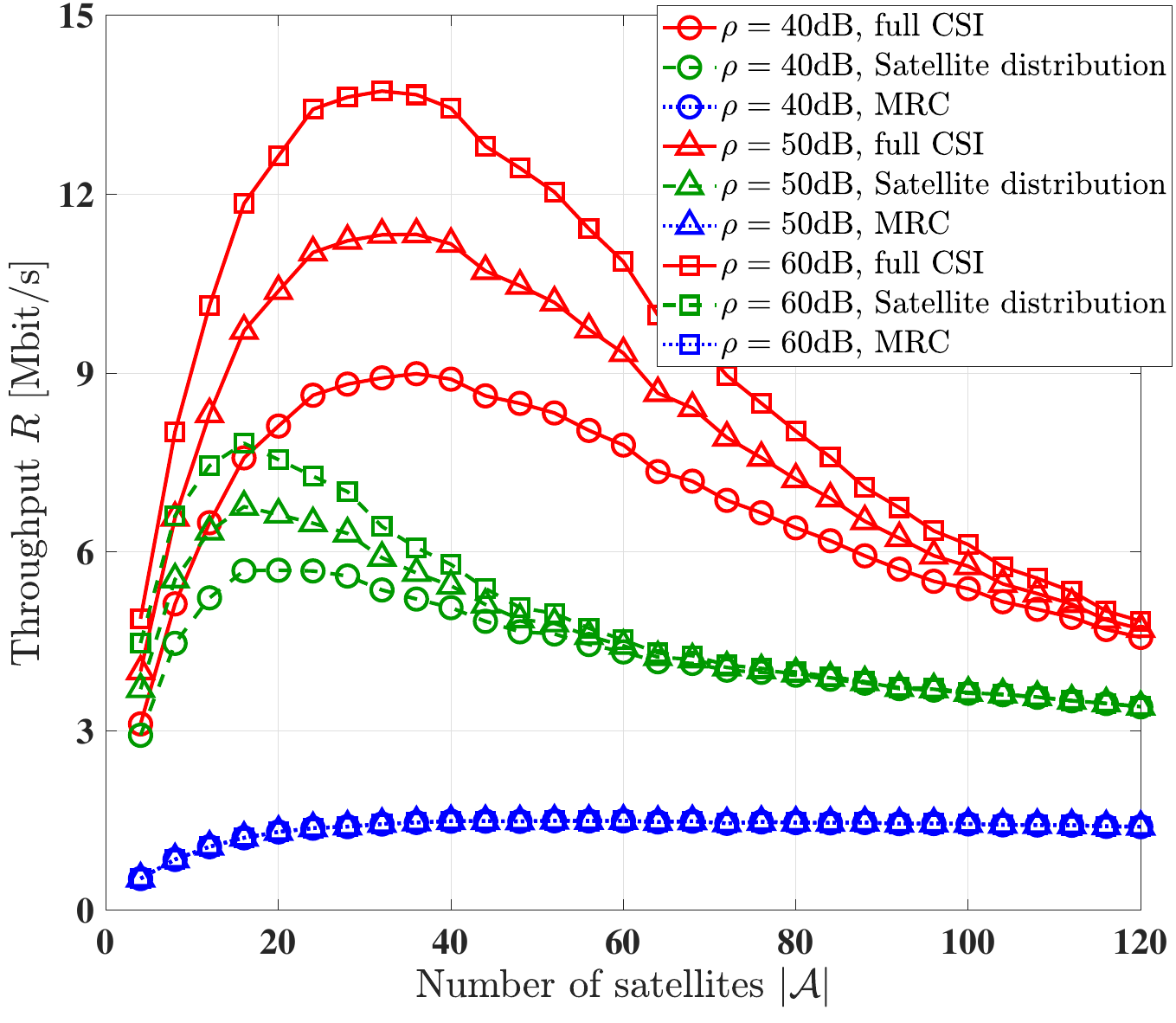}
    \caption{The throughput $R$ versus the number of satellites $|\mathcal{A}|$, at different transmit power $\rho$.}\label{Simu_figure_10}
\end{figure}

In Fig.~\ref{Simu_figure_10}, the throughput $R$ is compared against the number of satellites $|\mathcal{A}|$ for different transmit power levels $\rho$. When the satellites are sparse, i.e., $|\mathcal{A}|<30$, the throughput can be improved by increasing the number of satellites. This can be attributed to the fact that when the satellites are sparse, the throughput is primarily determined by the probability $P_\mathrm{S}$ of the serving satellite located in the visible domain. Increasing the number of satellites enhances this probability $P_\mathrm{S}$, thereby improving the throughput. However, as the number of satellites further increases, the throughput will be degraded due to the increased inter-satellite interference.

\begin{figure}[!t]
    \centering
    \includegraphics[width=2.35in]{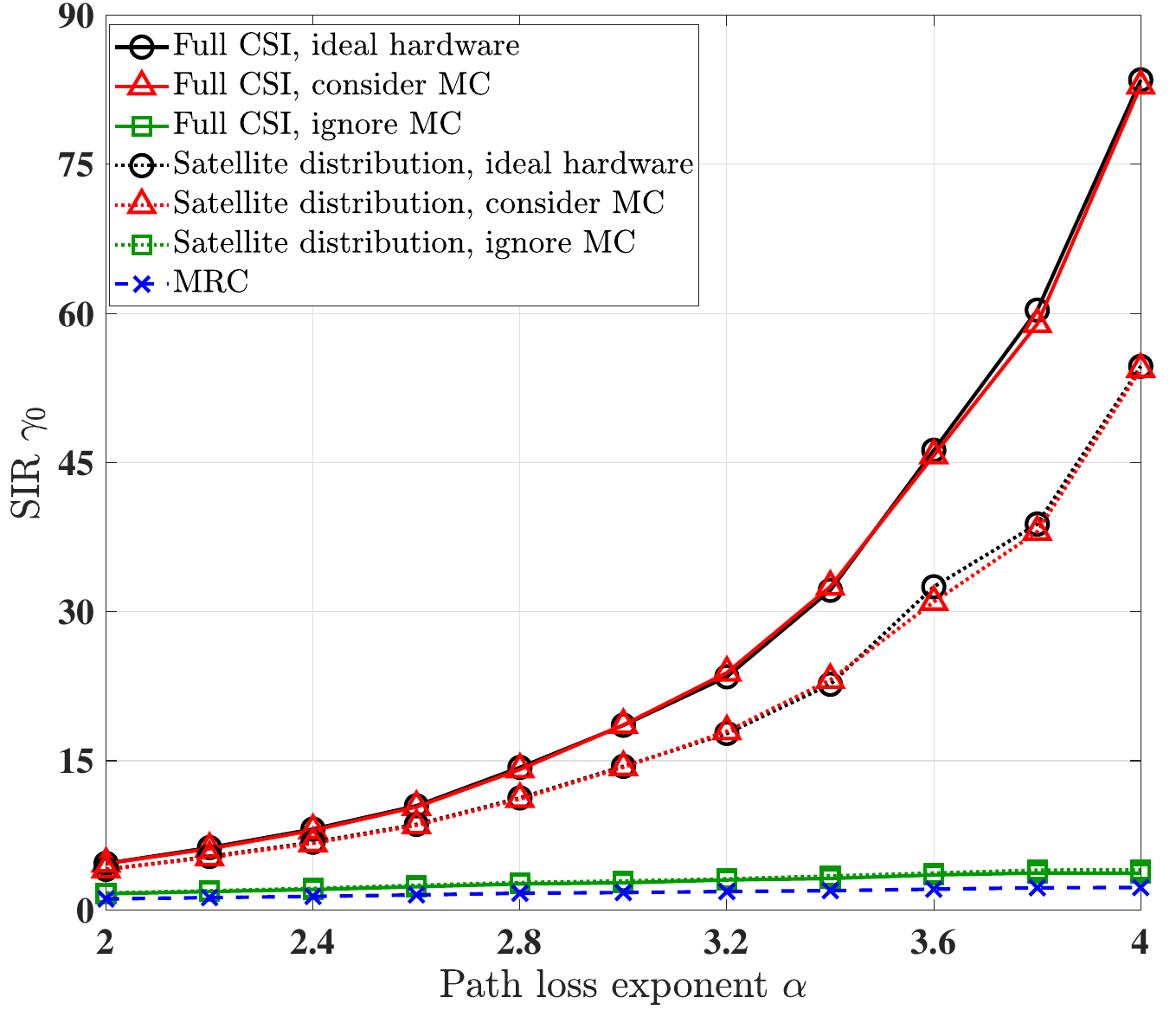}
    \caption{Comparison of the SIR $\gamma_0$ versus the path loss exponent $\alpha$ in various beamforming methods.}\label{Simu_figure_11}
\end{figure}

In satellite communications, the path loss exponent varies due to factors like atmospheric conditions, terrain features, and ground reflections within the propagation environment. Fig.~\ref{Simu_figure_11} compares the signal-to-interference ratio (SIR) denoted as $\gamma_0$, versus the path loss exponent $\alpha$ for the various beamforming methods. This shows that the throughput can be improved as the path loss exponent $\alpha$ increases, since a higher path loss exponent mitigates inter-satellite interference.

\begin{figure}[!t]
    \centering
    \includegraphics[width=2.35in]{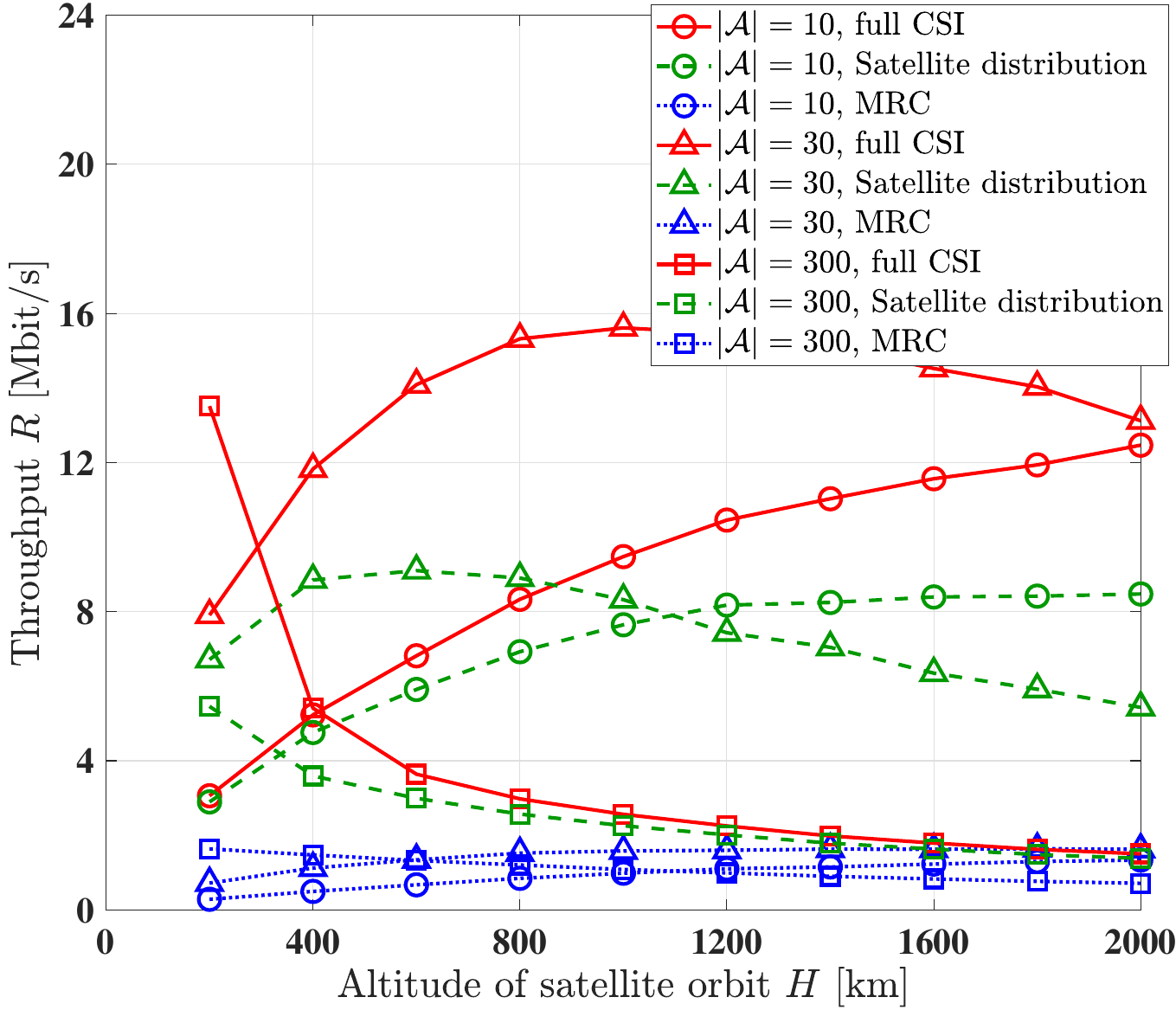}
    \caption{The throughput $R$ versus the altitude of satellite orbit $H$, for 10, 30 and 300 satellites.}\label{Simu_figure_12}
\end{figure}

Finally, to investigate the effect of varying the altitude of the satellite constellation on the throughput, we consider the specific case of a single altitude, i.e., when $H_1=H_2=H$. Fig.~\ref{Simu_figure_12} compares the throughput $R$ versus the altitude of the satellite orbit $H$ for different numbers of satellites $|\mathcal{A}|$. When the number of satellites is small, i.e. $|\mathcal{A}|=10$, a higher throughput can be attained as the altitude of the satellite constellation increases. This can be explained by the fact that when the satellite density is low, increasing the altitude improves signal coverage probability. However, when the number of satellites is moderate, i.e. $|\mathcal{A}|=30$, the throughput initially increases and then decreases with the altitude of the satellite constellation. Conversely, when the number of satellites is high, i.e. $|\mathcal{A}|=300$, the throughput consistently decreases as the altitude of the satellite constellation increases. This can be attributed to the fact that at high satellite densities, increasing the altitude exacerbates inter-satellite interference.

\section{Conclusions}\label{Conclusion}
In this paper, we conceived a hybrid beamforming design for holographic metasurface-based multi-altitude LEO satellite networks. Specifically, the holographic beamformer was optimized for maximizing the channel gain of the serving satellite link, while the digital beamformer was designed for mitigating the interference. To reduce the CSI acquisition overhead, we proposed the low-complexity MMSE RC scheme based on the statistical information of the LEO satellite constellation by leveraging stochastic geometry, which achieves comparable throughput to using full CSI in the case of a dense deployment of LEO satellites. Furthermore, the holographic metasurface-based hybrid beamformer can achieve higher throughput than the full digital beamformer relying on the SoA antenna array, given the same physical size of the transceivers. The dense placement of holographic metasurface elements underscores the importance of explicitly considering the mutual coupling in the beamforming design.

\appendices
\section{Proof of \textit{Theorem} \ref{Theorem_1}}\label{Appendix_A}
Firstly, we consider the case of $\sqrt{H_1(2R_\mathrm{e}+H_1)}\leq H_2$. We define the ball $B(d)=\{x^2+y^2+(z-R_\mathrm{e})^2\leq d^2|(x,y,z)\in\mathbb{R}^3\}$, which represents the set of points whose distance from the terrestrial user is no higher than $d$. When $d\in[0,H_1)$, the volume of the intersection of the ball $B(d)$ and the visible area $\Omega'$, denoted as $V_{B\cap\Omega'}(d)$, is
\begin{align}\label{Appendix_A_1}
    V_{B\cap\Omega'}(d)=0.
\end{align}
When $d\in[H_1,\tilde{H}_1)$, the volume of the intersection of the ball $B(d)$ and the visible domain $\Omega'$ is
\begin{align}\label{Appendix_A_2}
    \notag V_{B\cap\Omega'}(d)=&\frac{\pi}{4R_\mathrm{e}}d^4+\frac{2\pi}{3}d^3
    -\frac{\pi H_1\left(2R_\mathrm{e}+H_1\right)}{2R_\mathrm{e}}d^2\\
    &+\frac{\pi H_1^3\left(4R_\mathrm{e}+3H_1\right)}{12R_\mathrm{e}}.
\end{align}
By contrast, when $d\in[\sqrt{H_1(2R_\mathrm{e}+H_1)},H_2)$, the volume of the intersection of the ball $B(d)$ and the visible shell $\Omega'$ is
\begin{align}\label{Appendix_A_3}
    V_{B\cap\Omega'}(d)=\frac{2\pi}{3}d^3-\frac{\pi H_1^2\left(3R_\mathrm{e}+2H_1\right)}{3}.
\end{align}
For $d\in[H_2,\sqrt{H_2(2R_\mathrm{e}+H_2)})$, the volume of the intersection of the ball $B(d)$ and the visible region $\Omega'$ is
\begin{align}\label{Appendix_A_4}
    \notag V_{B\cap\Omega'}(d)=&-\frac{\pi}{4R_\mathrm{e}}d^4
    +\frac{\pi H_2\left(2R_\mathrm{e}+H_2\right)}{2R_\mathrm{e}}d^2\\
    &-\frac{\pi\left(H_2^3\left(4R_\mathrm{e}+3H_2\right)+
    4R_\mathrm{e}H_1^2\left(3R_\mathrm{e}+2H_1\right)\right)}{12R_\mathrm{e}}.
\end{align}
Finally, when $d\in[\sqrt{H_2(2R_\mathrm{e}+H_2)},\infty)$, the volume of the intersection of the ball $B(d)$ and the visible region $\Omega'$ is
\begin{align}\label{Appendix_A_5}
    V_{B\cap\Omega'}(d)=V'.
\end{align}
According to (\ref{Appendix_A_1}), (\ref{Appendix_A_2}), (\ref{Appendix_A_3}), (\ref{Appendix_A_4}) and (\ref{Appendix_A_5}), the CDF of the distance from any specific one of the satellite in the visible region $\Omega'$ to the terrestrial user is
\begin{align}\label{Appendix_A_6}
    F_{D'}(d)=\frac{V_{B\cap\Omega'}(d)}{V'},
\end{align}
as shown in (\ref{System_Model_9}).

By contrast, when $\sqrt{H_1(2R_\mathrm{e}+H_1)}>H_2$, the CDF of the distance from any specific satellite in the visible region to the terrestrial user can be derived as shown in (\ref{System_Model_10}).

\section{Proof of \textit{Theorem} \ref{Theorem_2}}\label{Appendix_B}
Given the distance between the serving satellite and the terrestrial user $D_0=d_0$, the value of $f_{D_\mathrm{I}'|D_0}(d|d_0)$ can be derived as
\begin{align}\label{Appendix_B_1}
    f_{D_\mathrm{I}'|D_0}(d|d_0)=\frac{f_{D'}(d)}{1-F_{D'}(d_0)}.
\end{align}
Substituting (\ref{System_Model_11}) and (\ref{System_Model_12}) into (\ref{Appendix_B_1}), we can arrive at $f_{D_\mathrm{I}'|D_0}(d|d_0)$ as shown in (\ref{System_Model_15}) and (\ref{System_Model_16}).

\section{Proof of \textit{Theorem} \ref{Theorem_3}}\label{Appendix_C}
According to (\ref{System_Model_2}), (\ref{System_Model_3}), (\ref{System_Model_4}) and (\ref{Beamforming_Design_14}), $\mathbb{E}[\mathbf{h}^{(l)}\mathbf{h}^{(l)\mathrm{H}}]$ can be expressed as
\begin{align}\label{Appendix_C_1}
    \notag\mathbb{E}\left[\mathbf{h}^{(l)}\mathbf{h}^{(l)\mathrm{H}}\right]
    =&\varsigma\zeta\left(\frac{\lambda}{4\pi}\right)^2\mathcal{L}(d_0)\cdot\\
    &\frac{\mathbf{Q}\left(\mathbf{C}\mathbf{C}^\mathrm{H}
    +\left(\mathbf{C}\mathbf{C}^\mathrm{H}\right)\odot\mathbf{I}_{MN}\right)
    \mathbf{Q}^\mathrm{H}}{4},
\end{align}
where $\mathcal{L}(d_0)=\mathbb{E}\left[d_0^{-\alpha}\right]$ can be further represented as shown in (\ref{Beamforming_Design_20}), (\ref{Beamforming_Design_21}), (\ref{Beamforming_Design_22}), (\ref{Beamforming_Design_23}), (\ref{Beamforming_Design_24}),(\ref{Beamforming_Design_25}), (\ref{Beamforming_Design_26}) and (\ref{Beamforming_Design_27}) by substituting $d=d_0$ into (\ref{System_Model_11}) and (\ref{System_Model_12}). Furthermore, the average number of interfering satellites located in the visible region, denoted as $|\mathcal{A'}|$, is
\begin{align}\label{Appendix_C_2}
    |\mathcal{A'}|=\left(|\mathcal{A}|-1\right)P_\mathrm{I}.
\end{align}
According to (\ref{Appendix_C_1}) and (\ref{Appendix_C_2}), we have:
\begin{align}\label{Appendix_C_3}
    \notag\mathbf{R}_\mathrm{I}'
    =&\mathbb{E}\left[\sum_{\mathbf{p}_l\in\Omega'}
    \mathbf{h}^{(l)}\mathbf{h}^{(l)\mathrm{H}}\right]\\
    \notag=&|\mathcal{A'}|\cdot\mathbb{E}\left[\mathbf{h}^{(l)}\mathbf{h}^{(l)\mathrm{H}}\right]\\
    \notag=&\varsigma\zeta\left(\frac{\lambda}{4\pi}\right)^2\left(|\mathcal{A}|-1\right)
    P_\mathrm{I}\mathcal{L}(d_0)\cdot\\
    &\frac{\mathbf{Q}\left(\mathbf{C}\mathbf{C}^\mathrm{H}
    +\left(\mathbf{C}\mathbf{C}^\mathrm{H}\right)
    \odot\mathbf{I}_{MN}\right)\mathbf{Q}^\mathrm{H}}{4}.
\end{align}

\bibliographystyle{IEEEtran}
\bibliography{IEEEabrv,TAMS}

\end{document}